\documentclass[reprint,twocolumn,superscriptaddress,preprintnumbers,amsmath,amssymb,aps,prd,floatfix]{revtex4-2}

\usepackage[utf8]{inputenc}

\usepackage{mathtools}
\usepackage{amsfonts}
\usepackage{mathrsfs}
\usepackage{bm}
\usepackage{bbm}
\usepackage[normalem]{ulem}
\usepackage{bbold}
\usepackage{physics}
\usepackage{braket}
\usepackage{slashed}
\usepackage{tensor}
\usepackage{dsfont}
\usepackage{float}
\usepackage{dcolumn}
\usepackage{amssymb,amsmath}
\usepackage{titlesec}

\usepackage{graphicx}
\usepackage{subcaption}
\usepackage{color}
\usepackage{array}

\usepackage{placeins}
\usepackage{booktabs}

\usepackage{xspace}
\usepackage{hyperref}
\usepackage[nameinlink]{cleveref}
\usepackage{bookmark}
\usepackage{units}
\hypersetup{
		colorlinks,
		linkcolor={red!75!black},
		citecolor={blue!75!black},
		urlcolor={blue!75!black}
	}

\usepackage{booktabs}
\usepackage{multirow}

\newcolumntype{C}{>{$}c<{$}}
\AtBeginDocument{
		\heavyrulewidth=.08em
		\lightrulewidth=.05em
		\cmidrulewidth=.03em
		\belowrulesep=.65ex
		\belowbottomsep=0pt
		\aboverulesep=.4ex
		\abovetopsep=0pt
		\cmidrulesep=\doublerulesep
		\cmidrulekern=.5em
		\defaultaddspace=.5em
	}

\setkeys{Gin}{width=0.48\textwidth}
\usepackage{xifthen}
\usepackage{xcolor}

\usepackage{booktabs}
\usepackage{multirow}

\graphicspath{{./figures}}

\newcommand{\orcid}[1]{\href{https://orcid.org/#1}{\includegraphics[height=1.9ex,width=1.9ex]{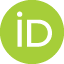}}}

\newcommand{\gettitle}{Solving sign problems with physics-informed kernels}

\newcommand{\getHeidelbergAffiliation}{\affiliation{Institut f{\"u}r Theoretische Physik, Universit{\"a}t Heidelberg, Philosophenweg 16, 69120 Heidelberg, Germany}}

\newcommand{\getBochumAffiliation}{\affiliation{Theoretische Physik III, Ruhr-Universit{\"a}t Bochum, Universit{\"a}tsstraße 150, 44721 Bochum, Germany}}

\newcommand{\getEMMIAffiliation}{\affiliation{ExtreMe Matter Institute EMMI, GSI, Planckstr. 1, 64291 Darmstadt, Germany}}

\hypersetup{
	pdftitle={\gettitle},
	pdfauthor={Ihssen, Kapust, Pawlowski},
	pdfkeywords={} ,
	bookmarksopen=true,
	bookmarksopenlevel=2,
	bookmarksnumbered=true
}

\begin{document}
	
\title{\gettitle}
	\author{Friederike Ihssen \orcid{0000-0002-1550-3423}\,}
	\getHeidelbergAffiliation
	\getBochumAffiliation
	\author{Renzo Kapust \orcid{0009-0009-1163-6315}\,}
	\getHeidelbergAffiliation
	\author{Jan M. Pawlowski \orcid{0000-0003-0003-7180}\,}
	\getHeidelbergAffiliation
	\getEMMIAffiliation

\begin{abstract}
 In the present work we construct a novel generative architecture for systems with complex probability distributions. In general, these sampling tasks come with two challenges: resolving sign problems and efficient sampling. The architecture is based on  \textit{physics-informed kernels} (PIKs) introduced in \cite{Ihssen:2025ybn}, and aims at resolving both challenges. Key to the complex PIK-architecture is its probability-weight preserving property, which allows us to map the sampling task to one on a sign-problem free manifold with a simple distribution and efficient sampling. The potential of this novel architecture is demonstrated within applications to zero-dimensional field theories with complex couplings, as well as the real-time evolution of the quantum-mechanical harmonic oscillator. 
\end{abstract}

\maketitle

\textit{Introduction.---} 
%
Many interesting physics systems come with complex `distributions' and `weights' and exhibit sign problems that prohibit the simple application of standard Monte Carlo sampling.  
Chiefly important examples include QCD at finite density, fermionic systems in statistical and condensed matter physics with spin or mass imbalance and real-time quantum systems. 

In the present work we propose a novel sampling architecture based on \textit{physics-informed kernels} (PIKs) introduced in \cite{Ihssen:2025ybn} for the sampling of lattice field theories and beyond. One of its core properties is the analytic access to the information transport or reparametrisation of the theory, encoded in the PIK. This allows us to map the sampling task of the underlying system with an oscillatory distribution to one with a real and positive distribution or with a slowly varying phase. In short, the  PIK-architecture is based on \textit{weight-preserving} deformations of the complex distribution and the sampling manifold of the system at hand. Importantly, the intricacies known from other related approaches such as boundary terms and run-away trajectories in CLE~\cite{Aarts:2009uq}, or the classification and summation over all relevant thimbles in the Lefschetz thimble approach~\cite{Tanizaki:2015rda,Zambello:2019lva}, are absent. These key properties are illustrated comprehensively within zero-dimensional models with analytic control as well as the real-time evolution of quantum-mechanical systems.\\[-2ex]

\textit{Complex sampling and the sign problem.---} 
%
PIK-architectures \cite{Ihssen:2025ybn} have been mainly developed for the efficient sampling with normalised distributions of lattice field theories. In the present work we consider a scalar field theory with a real scalar field and a generic complex distribution for illustrative purposes. We define 
\begin{subequations} 
	\label{eq:ExampleTheory}
\begin{align} 
	p(\varphi) = e^{-S(\varphi)} \in\mathbbm{C}\,,\qquad S(\varphi) = \hat S(\varphi)+ \log {\cal N}\,. 
\label{eq:pvarphi}
\end{align}
Here, $S(\varphi)\in \mathbbm{C}$ is the action of the complex distribution $p(\varphi)\in\mathbbm{C}$, with $ \hat S(\varphi=0)=0$. The constant term $\log {\cal N}\in \mathbbm{C}$ is nothing but the normalisation of the distribution $p(\varphi)$ which guarantees 
\begin{align} 
Z= 	 \int\limits_{{\cal M}} \! d\varphi  \, p(\varphi)=1\,, \qquad \int\limits_{{\cal M}} \!d\varphi = \prod_{ i} \int\limits_\mathbbm{R} d\varphi_{ i}\,. 
\label{eq:Normalisationpvarphi}
\end{align}
where $ i$ labels the data points.
In our lattice example, the \textit{real} field $\varphi$ is defined on a finite (hyper-)cubic lattice ${\cal D}$ with  
\begin{align} 
	\varphi: {\cal D}\to {\cal M}=\mathbbm{R}^{\cal D}\,,\qquad 	{\cal D} = [0,1,...,N_{\textrm{max}}]^d\,. 
	\label{eq:varphiMap}
\end{align} 
\end{subequations} 
We will not specify the action $\hat S(\varphi)$; typically  it includes a kinetic term (hopping term) and an on-site self-interaction (potential term). For such an example of a real scalar lattice field theory we refer the reader to \cite{Ihssen:2025ybn}. In the general setup, ${\cal D}$ labels a continuous or discrete data set with data $\varphi$, and the continuous or discrete sampling manifold ${\cal M}$ is generalised likewise. All these setups are defined by \textit{sampling pairs} $(p(\varphi)\,,{\cal M})$ of the probability distribution and the sampling manifold. 

In general, sampling with a complex distribution $p(\varphi)\in\mathbbm{C}$ in \labelcref{eq:pvarphi} poses a numerical challenge, and for highly oscillatory distributions the sampling task may exhibit a sign problem. A rather generic and seemingly simple resolution of such a sign problem in our example \labelcref{eq:ExampleTheory} is provided by the deformation of the integration manifold ${\cal M}$ with a highly oscillatory distribution $p(\varphi)$ into an integration manifold ${\cal M}_1\subset  \mathbbm{C}^{\cal D}$, on which $p(\varphi)$ exhibits no sign problem. 

Assuming that this deformation step has been performed, it leaves us with a secondary problem of efficiently sampling from the given distribution $p(\varphi)$. This secondary problem can be remedied by mapping $p(\varphi)$ to a simple distribution such as a normal distribution $p_0(\phi)$ where $\phi\in {\cal M}_0\subset  \mathbbm{C}^{\cal D}$. While normal distributions with the sampling manifold ${\cal M}_0 = {\cal M}$ are standard choices, the set of viable $(p_0,{\cal M}_0)$ is not restricted to them. The only requirement is efficient sampling while keeping all symmetries. \\[-2ex]

\textit{Complex sampling with physics-informed kernels.---} 
%
In the present work we use the PIK-architecture \cite{Ihssen:2025ybn} to resolve the combined sign problem and efficient sampling tasks discussed above: in general, the physics-informed kernel (PIK) provides a map from some sampling pair $( p_0(\phi)\,,\, {\cal M}_0)$ to that under investigation, 
\begin{align} 
	\left( p_0(\phi)\,,\, {\cal M}_0\right)\longrightarrow 	\left(p_1(\phi)\,,\, {\cal M}_1\right)\,, \qquad  p_1(\phi) =p(\phi)\,.  
\label{eq:GlobalMap}
\end{align} 
Consequently, the PIK-architecture allows for an efficient sampling of the pair $(p(\phi)\,,{\cal M}_1)$, if the original pair can be sampled efficiently. This solves the given sampling task if the sampling on ${\cal M}_1$ is equivalent to that on the original sampling manifold ${\cal M}$, 
\begin{align} 
\int\limits_{{\cal M}_1}  d\mu_1 \, p(\phi)\, {\cal O}(\phi)\stackrel{!}{=} \int\limits_{\cal M}  d\mu \,p(\varphi)\,{\cal O}(\varphi)\,, 
	\label{eq:EquivalenceM}
\end{align}  
for general operators ${\cal O}(\phi)$. Here, $d\mu_1(\phi)$ and $d\mu(\varphi)$ are the volume elements of ${\cal M}_1$ and ${\cal M}$, embedded in the complexification ${\cal C}$ of ${\cal M}$ and $\varphi$ denotes the variable in the original complex sampling problem. \Cref{eq:EquivalenceM} is satisfied if ${\cal M}_1$ can smoothly be deformed into ${\cal M}$ without swiping over singular points in the distribution $p(\phi)$ and the operators ${\cal O}(\phi)$. 

In our lattice example the sampling manifold is given by ${\cal M}= \mathbbm{R}^{\cal D}$ and its complexification is given by ${\cal C}= \mathbbm{C}^{\cal D}$.  For the sake of simplicity we will consider this explicit example from now on without loss of generality. 

The key property of the PIK-map in \labelcref{eq:GlobalMap} is that it is weight-preserving. This entails in particular, that the sampling with $( p_1(\phi)\,,\, {\cal M}_1)$ is real, \textit{if} the sampling pair $( p_0(\phi)\,,\, {\cal M}_0)$ is real. More generally, if the, potentially complex, sampling pair $( p_0(\phi)\,,\, {\cal M}_0)$ can be sampled efficiently and has no sign problem, it induces an efficient and sign- and overlap-problem free sampling for the sampling pair $( p_1(\phi)\,,\, {\cal M}_1)$. 

We proceed by discussing the computation of the physics-informed kernel (PIK), that is the differential map $\dot\phi(\phi)$. In the PIK-architecture \cite{Ihssen:2025ybn} the local deformation of the sampling manifold and the distribution is governed by the Wegner equation \cite{Wegner_1974}, 
\begin{align}
	\frac{d \,p_t(\phi)}{d t} = -\frac{\partial}{\partial\phi_i }\Bigl[ \dot\phi_{t,i}(\phi) \, p_t(\phi)\Bigr] \,,\qquad t\in [0,1]\,,
	\label{eq:WegnerEquation} 
\end{align}
with the corresponding PIK $\dot\phi_t(\phi)$ of the deformation, and the RG-time $t$.  The index $i\in{\cal D}$ in \labelcref{eq:WegnerEquation} is summed over the data points. \Cref{eq:WegnerEquation} depends on a $t$-dependent normalised distribution  
\begin{align}
	p_t (\phi) = e^{- S_t(\phi)}\,, 
	\label{eq:Defofpt}
\end{align}
with $S_t = \hat S_t + \log \mathcal{N}_t$.
Importantly, $\hat S_t$ is chosen analytically to interpolate between $p_0(\phi)$ and $p_1(\phi)$.
In our lattice example $\phi$ is the complex field variable on the lattice ${\cal D}$, 
\begin{align} 
	\phi: {\cal D}\to \mathbbm{C}^{\cal D}\,. 
	\label{eq:phiMap}
\end{align} 
In general, ${\cal D}$  is the set of data points and $\phi$ takes values in the complexification of the data range. 

The complexification of $\phi$ has two sources. The first rather trivial one is the possible choice of a complex initial manifold ${\cal M}_0\subset \mathbbm{C}^{\cal D}$. The second source is the deformation $\partial_t{\cal M}_t$ of the manifold ${\cal M}_t$ due to the complex transformation $\dot\phi_t$ itself. With \labelcref{eq:Defofpt}, we rewrite \labelcref{eq:WegnerEquation} as a differential equation for the physics-informed RG-pair (PIRG pair) $(S_t(\phi)\,,\,\dot\phi)$, 
\begin{align}
	\frac{d\,S_t(\phi)}{dt} +\dot{\phi}_t (\phi) \,\frac{\partial S_t(\phi)}{\partial \phi}= \frac{\partial  \dot{\phi}_t(\phi)}{\partial \phi} \,, 
	\label{eq:WF-dotphi}
\end{align}
where we have suppressed the indices $i\in{\cal D}$ for the sake of readability. 
For complex $d S_t(\phi)/dt $ and/or $\partial S_t(\phi)/\partial\phi$, the map $\dot\phi(\phi)$ is complex and so is the differential manifold $\partial_t{\cal M}_t$. Note that for smooth $\dot\phi(\phi)$, the manifold $\partial_t {\cal M}_t\subset\mathbbm{C}^{\cal D}$ is a closed or open hyper-surface, an example with one complex dimension and further explanations can be found in the supplement. 

Accordingly,  
\labelcref{eq:WegnerEquation,eq:WF-dotphi} comprise a general differential reparametrisation of the path integral with a smooth deformation of the integration manifold with the PIK $\dot\phi(\phi)\in \mathbbm{C}^{\cal D}$, if the latter is invertible and non-singular. 

Now we set up the complex PIK-architecture for the sign-problem free sampling task: With \labelcref{eq:WegnerEquation,eq:WF-dotphi} we map the sampling with $p(\varphi)$ on ${\cal M}$ with a sign problem to the efficient sampling with $p_0(\phi)$ on the sign-problem free manifold ${\cal M}_0$. We find for all $t$,  
\begin{subequations} 
\label{eq:ComplexPIKwoSignProblem}
\begin{align} 
	\int\limits_{{\cal M}_t}   d\mu_t  \, p_t(\phi) \,{\cal O}(\phi)= \int\limits_{{\cal M}_0}  d\mu_0  \,p_0(\phi)\,{\cal O}\bigl(\phi_t(\phi)\bigr)\,, 
\label{eq:EquivalenceSampling1}
\end{align} 
where 
\begin{align} 
\phi_t(\phi) = \phi + \int_0^{t} ds \, \dot \phi_{s} (\phi_{s})\,, \qquad \varphi(\phi)=\phi_1(\phi)\,,
	\label{eq:IntegrateMap}
\end{align}
\end{subequations} 
and $d \mu_t$ denotes the volume element of the corresponding manifold ${\cal M}_t$.
\Cref{eq:ComplexPIKwoSignProblem} with \labelcref{eq:EquivalenceM}, evaluated for $t=1$, leads to the aimed for equivalence 
\begin{align} 
	\int\limits_{{\cal M}}  d\mu \, p(\varphi) \,{\cal O}(\varphi)= \int\limits_{{\cal M}_0}  d\mu_0  \,p_0(\phi)\,{\cal O}\bigl(\varphi(\phi)\bigr)\,. 
	\label{eq:EquivalenceSampling}
\end{align} 
The existence of the global map \labelcref{eq:IntegrateMap} was discussed in detail in \cite{Ihssen:2025ybn} for real actions $S_t$ and the existence criteria apply to complex $S_t$ without modification. The proof of \labelcref{eq:EquivalenceSampling} as well as further details can be found in the supplement. Here we only emphasise that the physics or more generally the complete information of the underlying distribution $p(\varphi)$ is carried by the map $\varphi(\phi)$ alone. The analytic access to this map, and the factorisation of the global computing task into the independent ones of computing $\dot\phi_t(\phi)$ for all $t\in[0,1]$ are amongst the advantageous properties of the PIK-architecture, see again \cite{Ihssen:2024ihp,Ihssen:2025ybn}. 

It is left to prove that the right-hand side of \labelcref{eq:EquivalenceSampling} is sign-problem free. For that purpose we show, that the sampling with $p_t(\phi)$ on ${\cal M}_t$ has no sign and overlap problems, if the sampling with $(p_0(\phi)\,,\,{\cal M}_0 )$ has no sign problem. This important property can be deduced from the \textit{weight-preserving} property of the PIK-architecture. For that purpose we shall prove that the flow of the 'statistical' weight of general intervals ${\cal T}_t$ vanishes, that is 
\begin{align} 
	\frac{d\,P_t({\cal T}_t)}{dt}  = 0\,,\qquad P_t({\cal T}_t)  = \int_{{\cal T}_t}   d\mu_t \, p_t(\phi) \,.
	\label{eq:dtWeightPT=0}
\end{align}
Here, ${\cal T}_t$ is the image of a given interval ${\cal T}_0$ with the map $\phi_t(\phi)$ in \labelcref{eq:IntegrateMap}. A direct proof of \labelcref{eq:dtWeightPT=0} with \labelcref{eq:WegnerEquation} is provided in the supplement. Here, we simply consider the integral of \labelcref{eq:dtWeightPT=0} from the initial RG-time to $t$, which is $P_t({\cal T}_t)=P_0({\cal T}_0)$. This identity can be derived from \labelcref{eq:ComplexPIKwoSignProblem} by using the characteristic function $ \chi^{\ }_{{\cal T}_t}$ of the sub-manifold ${\cal T}_t$ as the operator ${\cal O}$. Then, $\chi^{\ }_{{\cal T}_t}(\phi_t(\phi))=\chi^{\ }_{{\cal T}_0}(\phi)$ is the characteristic function of ${\cal T}_0$ on ${\cal M}_0$. Hence, inserting this operator in \labelcref{eq:ComplexPIKwoSignProblem} leads us to
\begin{align} 
	\int\limits_{{\cal T}_t}   d\mu_t  \, p_t(\phi) = \int\limits_{{\cal T}_0}  d\mu_0  \,p_0(\phi)\,, 
	\label{eq:EquivalentWeight}
\end{align} 
and the $t$-derivative of \labelcref{eq:EquivalentWeight} leads us to \labelcref{eq:dtWeightPT=0}. If we use infinitesimal sub-manifolds 
${\cal T}_t$, we readily find that the differential distribution or weight $d\mu_t\, p_t(\phi)$ is equivalent to $d\mu_0\, p_0(\phi)$. The differential equivalence is elucidated with the pull-back of the left-hand side to the pre-image $\mathcal{T}_0$, to wit,  
\begin{align}
	\int\limits_{{\cal T}_0} \! d\mu _0\det\!\left[ \frac{\partial \phi_t }{\partial \phi_0} \right] \,e^{- S_t\bigl(\phi_t(\phi_0)\bigr)} \equiv \int\limits_{{\cal T}_0} \! d\mu_0\, e^{- S_0(\phi_0)} \,. 
	\label{eq:Idenity}
\end{align}
As \labelcref{eq:Idenity} holds true for general sub-manifolds ${\cal T}_0$, it has to hold true for the integrands, and we arrive at 
\begin{align}
	S_t\bigl(\phi_t(\phi_0)\bigr) -  \textrm{Tr} \log \left[ \frac{\partial \phi_t }{\partial \phi_0} (\phi_0)\right] = S_0(\phi_0) \,. 
	\label{eq:SignProblem}
\end{align}
In short, \labelcref{eq:dtWeightPT=0} or \labelcref{eq:EquivalentWeight} entail that the map $\dot\phi$ is weight-preserving by construction and hence no overlap problem and sign problem is generated by the flow with $\dot\phi$. Hence, if the sampling pair at $t=0$ is chosen sign-problem free, so are all sampling pairs. In particular, if the sampling pair at $t=0$ is real, all sampling pairs are real, and specifically $(p_1(\phi)\,,\,{\cal M}_1)$.  We call the set  of weight-preserving manifolds 
\begin{align} 
	{\cal P}={\cal P}_1\,,\qquad {\cal P}_t=\{{\cal M}_s\,|\,s\in[0,t]\}\,,
	\label{eq:PIKfold}
\end{align} 
endowed with the holomorphic map $\varphi(\phi)$ on ${\cal P}$, the \textit{PIKfold} of the set of  flowing distributions $\{p_t\}$. The sets ${\cal P}_t$ are sub-PIKfolds. More details are provided in the supplement. 
Sign-problem free PIK-architectures are obtained by construction for a sign-problem free sampling pair  $(p_0, {\cal M}_0)$. \\[-2ex]

\textit{Solving sign problems with PIKs.---} 
%
The remaining task of solving sign problems with the PIK-architecture is that of finding a PIKfold. In many cases this task is completed by simply choosing a path $S_t$ from an initial action $S_0$ to the physical one with $S_1=S$. Then, if the PIK $\dot \phi$ can be integrated from $t=0$ to $t=1$, the PIKfold exists. Indeed, this is the case for all examples considered here. We emphasise, however, that it is  not possible to integrate the differential map $\dot \phi$ from $t=0$ to $t=1$ for \textit{all} paths $\{ p_t \}$ with a given initial sampling pair $(p_0, {\cal M}_0)$. Moreover, for a large set of initial sampling pairs $(p_0, {\cal M}_0)$, PIKfolds may not exist at all. 

Importantly, the \textit{global} existence of the map can be inferred \textit{locally} from $S_t$: all manifolds ${\cal M}_t$ can be constructed from $S_t(\phi)$ alone. We first restrict ourselves to real sampling pairs $(p_0, {\cal M}_0)$. Then we use that $p_t\, d \mu_t$ is the likelihood on the manifold ${\cal M}_t$ and we find with  \labelcref{eq:dtWeightPT=0}, 
\begin{align} 
\left[  d\phi_{R} + i\,d\phi_I \right] \left[  \textrm{Re}(p_t(\phi)) + i\textrm{Im}(p_t(\phi)) \right] \in  \mathbbm{R}\,.	
\label{eq:ConstructionMt}
\end{align} 	
Here $\phi_R(\sigma),\phi_I(\sigma)$ are the real and imaginary coordinates of the $2{\cal D}$-dimensional complexification of the ${\cal D}$-dimensional manifold ${\cal M}$, with some ${\cal D}$-dimensional coordinates $\sigma$ on the manifold ${\cal M}$. \Cref{eq:ConstructionMt} entails that the imaginary part of $p_t\, d \mu_t$ vanishes. This allows the construction of ${\cal M}_t$ from  $S_t$ via the computation of $(\phi_R(\sigma), \phi_I(\sigma))$.  

\begin{figure}[t]
	\centering
	\includegraphics[width=0.7\linewidth]{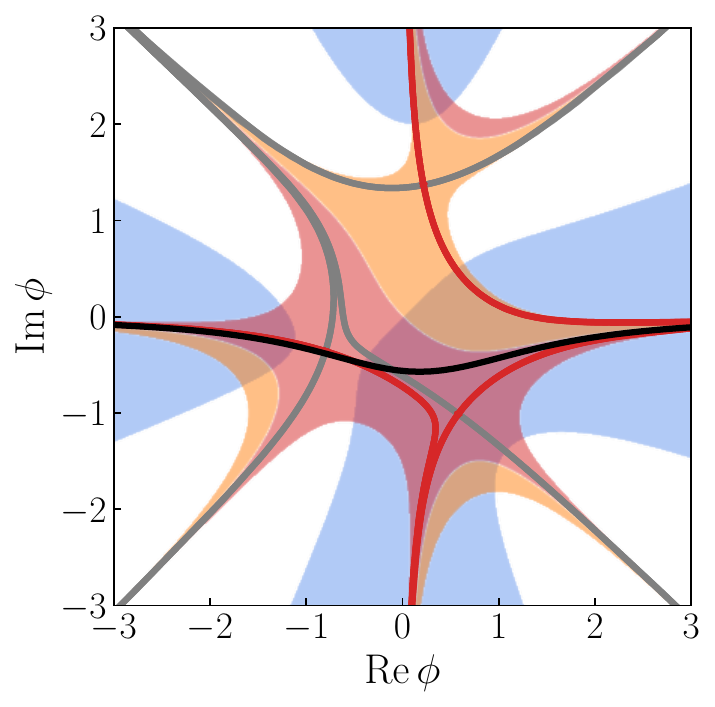}
	\caption{Final sampling manifold $\mathcal{M}_1$ \textit{(black)} for the zero-dimensional target action \labelcref{eq:Action0d}. The thimbles and anti-thimbles are shown in red and grey, respectively. The shaded areas indicate: $\Re[S(\phi)]>0$ \textit{(blue)} and $0 \leq \pm \Im[S(\phi)] \leq \frac{\pi}{2}$ \textit{(orange/red)}.\hspace*{\fill}}
	\label{fig:Showcase}
\end{figure}

We illustrate this property with a discussion of one-dimensional sampling tasks with ${\cal M}={\cal M}_0=\mathbbm{R}$ and a manifold ${\cal M}_t$ which is a monotonic function $\phi_R(\sigma)$ that covers $\mathbbm{R}$. Then, a vanishing imaginary part of \labelcref{eq:ConstructionMt} implies  
\begin{align} 
	\frac{d\phi_I}{d \phi_R} =-\frac{\textrm{Im}\left[p_t(\phi)\right]}{\textrm{Re}\left[p_t(\phi)\right]} = \tan \bigl( \Im[S_t(\phi_R + i \phi_I)]\bigr)\,. 
	\label{eq:DefofMt}
\end{align} 	
The solution $\phi_I(\phi_R)$ of the one-dimensional linear ordinary differential equation \labelcref{eq:DefofMt} defines the manifold ${\cal M}_t$ in the local coordinate system of $\phi$ in $t$. In \Cref{fig:Showcase} we depict the sampling manifold ${\cal M}_1$ (\textit{black line}) for the action \labelcref{eq:Action0d}. This example and its properties will be detailed further below. The potential absence of PIKfolds for \textit{any} real initial sampling pairs can also be inferred by solving \labelcref{eq:DefofMt} for $t=1$: if \labelcref{eq:DefofMt} admits no solution for $t=1$, there is no real PIKfold.   

Importantly, the computation of the PIKfold $\{{\cal M}_t\}$ with \labelcref{eq:DefofMt}, or more generally \labelcref{eq:ConstructionMt} and its complex generalisations, is not based on the evolution of the manifold from ${\cal M}_0$. The computation of the latter requires the solution of $\dot\phi_s$ between $0$ and $t$. Hence, \labelcref{eq:ConstructionMt} allows the classification of sign problems: if \labelcref{eq:ConstructionMt} can be solved for $t=1$, it admits real PIKfolds. In turn, if it cannot be solved for $t=1$, no real PIKfold exists. In the latter case one has to  start from complex sampling pairs $(p_0, {\cal M}_0)$ or one defines sampling manifolds beyond the PIKfold. This and further properties and generalisations are discussed in the supplement. These considerations are chiefly important for the application of the PIK-architecture to fermionic theories and theories with a finite chemical potential. However, it goes beyond the scope of the present work which is concerned with the construction of the novel PIK-architecture and will be discussed in \cite{PIKF2026}. \\[-2ex]

\textit{Solving sign problems in single-site models.---} 
%
The $\phi^4$-theory on a single site allows for important first benchmarks of approaches to the sign problem. As a comprehensive example we consider a $\phi^4$-theory with an additional complex linear term, 
\begin{align}
 \hat{S}_t(\phi) = \frac12 \phi^2  +  \frac14\phi^4  + (1 + i t) \phi \,. 
\label{eq:Action0d}
\end{align}
This example features both a global sign problem and a residual sign problem on each thimble~\cite{Aarts:2014nxa}. 
Further parameter choices for the single-site $\phi^4$-theory can be found in the supplement.

The initial sampling manifold is chosen as ${\cal M}_0=\mathbbm{R}$. With this choice we start with a real initial sampling pair and hence have a real PIKfold. This setup allows for efficient sampling with standard MC methods. 

The RG-time evolution of samples is fully determined by the analytical RG-time dependence of the action \labelcref{eq:Action0d}: For their numerical evaluation the time evolution is discretised in $N_t$ time-steps, at which the Wegner equation \labelcref{eq:WF-dotphi} is solved to yield the kernel $\dot{\phi}_t$. The latter is then used together with \labelcref{eq:IntegrateMap} to compute the final map $\varphi(\phi)$. Further technical details of this procedure are presented in the supplement.

The sign-problem free manifold ${\cal M}_1$ is depicted in \textit{black} in \Cref{fig:Showcase} and is either given by $\varphi(\phi)$, obtained from \labelcref{eq:IntegrateMap} at $t=1$ or from \labelcref{eq:DefofMt}. We also indicate the \textit{blue} regions corresponding to the Stokes-wedges with $\Re[ \hat{S}(\phi)]>0$. For $\phi \to \infty$ the sampling manifold is located within this region and finally converges to the real axis.

\begin{figure}[t]
	\centering
	\includegraphics[width=0.875\linewidth]{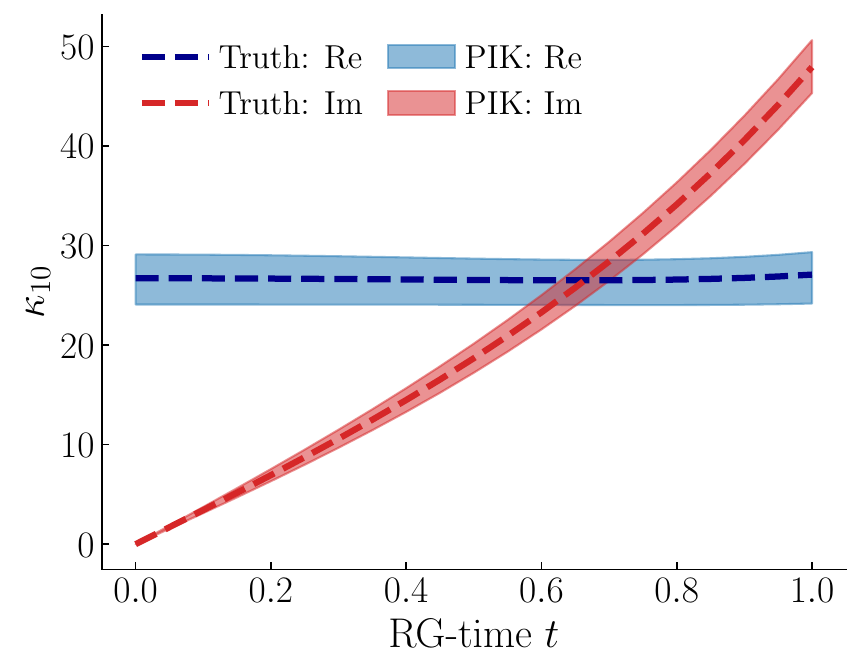} 
	\caption{RG-time dependent 10th cumulant of the model with the action \labelcref{eq:Action0d}. The cumulant is computed using $10^5$ Monte Carlo samples at $t=0$, which are subsequently transported to $t>0$ using \labelcref{eq:IntegrateMap} and the respective $\dot{\phi}$. As a visual guide we depict exact solutions computed by a direct numerical evaluation of the integral. \hspace*{\fill}}
	\label{fig:cumulant10}
\end{figure}

We also indicate the Lefschetz-thimble solutions, as well as \textit{orange} and \textit{red} regions corresponding to $0 \leq\pm \Im[ \hat{S}(\phi)]\, \leq \frac{\pi}{2}$, respectively. While there are two contributing thimbles, there is only one contour ${\cal M}_1$ with lightly varying phase in the action $S(\phi)$ but a real and positive $d\mu_t\,p_t(\phi)$ along the contour. This fact allows for efficient sampling on the entire (real) PIKfold ${\cal M}_t$ using \labelcref{eq:EquivalenceSampling} and we can follow the flow of all observables from their initial values in a real $\phi^4$-theory at $t=0$ to the physical ones at $t=1$. As a showcase, we depict the RG-time dependent 10th cumulant in \Cref{fig:cumulant10}.
The shown error corresponds to the statistical error of the MC simulation at $t=0$. Further details on the cumulants and their error sources can be found in the supplement. \\[-2ex]

%

\textit{Solving real-time sign problems.---} 
%
We now turn to simulations of field theories in a real-time setting. Specifically, we consider the quantum-mechanical harmonic oscillator. To construct a complex PIK in this setting we determine a path for the action $S_t$ that modifies its prefactor
\begin{align}\label{eq:prefactors}
	p_E(\varphi) \sim e^{- \hat{S}_E(\varphi)}  \qquad \to \qquad 	p_M(\varphi) \sim e^{i\, \hat{S}_M(\varphi)} \,,
\end{align}
where $E$ denotes Euclidean and $M$ Minkowski (space-) time. To accommodate \labelcref{eq:prefactors} we chose a path that rotates the Euclidean couplings into the complex plane
\begin{align}\label{eq:QMaction}
	\hat{S}_t(\phi) &= \sum_{i\in \mathcal{D}}  \frac{\kappa_t}{2}  (\phi_{i+1} - \phi_i)^2 -\left( m_t^2 + \epsilon\right)\frac{\phi_i^2}{2}  \,,
\end{align}
with the couplings given by
\begin{align}
	\kappa_t = e^{-i \frac{\pi}{2} t} \,,  \qquad	m_t^2 = m^2 e^{+i \frac{\pi}{2} t} \,,
\label{eq:RealTimePath}
\end{align}
and a small $\epsilon$, which ensures that the path integral is well-defined. In summary, this PIK-architecture uses samples of an Euclidean theory at $t=0$ with $\hat S_0 = \hat S_E$ for the computation of observables in the real-time theory at $t=1$ with $\hat S_1 = i \hat S_M$. The initial sampling pair is real and hence the setup has a sign- and overlap-problem free \textit{real} PIKfold. The kernel $\dot{\phi}$ is purely analytical, 
\begin{align}
	\dot{\phi}_t(\phi) = - \frac12 M_t^{-1} \frac{d  M_t }{dt}\, \phi \,,
\label{eq:freePIK1}
\end{align}
where the mass matrix $M_t$ is defined as $\hat{S}_t(\phi):= \frac12 \phi_i \, M_{t, i j}\phi_j$.
The full derivation can be found in the supplement. Importantly, it is a linear transformation of the fields, further highlighting the connection of PIKs to approaches such as the kernel-CLE \cite{Boguslavski:2022dee,Alvestad:2022abf,Alvestad:2023jgl}.

\begin{figure}[t]
	\centering
	\includegraphics[width=\linewidth]{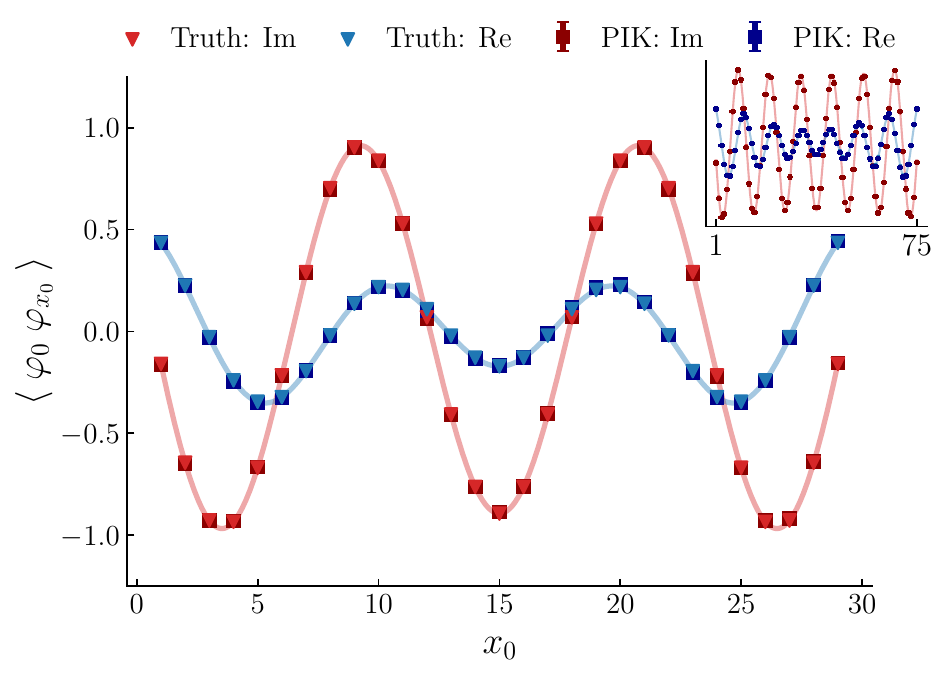}
	\caption{Two-point function $\langle \varphi_0\,\varphi_{x_0} \rangle$ of the real-time harmonic oscillator in 1+0 dimensions \labelcref{eq:QMaction}. The computation uses $N_{\textrm{max}}=30$ and $N_{\textrm{max}}=75$ (in-lay) lattice sites with couplings $m^2 = 0.3$, $\epsilon = \frac{2m}{N_{\textrm{max}}}$.\hspace*{\fill}}
	\label{fig:FreeTheoryCorrelations}
\end{figure}

In \Cref{fig:FreeTheoryCorrelations} we show the result for the two-point function $\langle \varphi_0 \, \varphi_{x_0} \rangle$ of the real-time theory for lattice sizes $N_{\textrm{max}}=30$ and $N_{\textrm{max}}=75$ in comparison to the analytic one. The computation was performed by sampling $10^5$ Euclidean configurations and the sampling with $p(\phi)$ was performed with the integrated PIK $\varphi(\phi)$ in 	\labelcref{eq:IntegrateMap}. 

Finally we remark that the key to the straightforward simulation of the real-time evolution of the harmonic oscillator is the existence of a \textit{real PIKfold}. Interestingly, \textit{real PIKfolds} also exist for the anharmonic oscillator with a few sites. It is suggestive that this important property persists for the full system, but a full discussion goes beyond the scope of the present work and shall be provided in  \cite{PIK-QM2026}. \\[-2ex]

%

\textit{Conclusions and outlook.---} 
%
We have set up a generative architecture for sampling tasks with complex distributions using the PIK approach put forward in \cite{Ihssen:2025ybn}. Key to the approach is the weight-preserving property of the complex PIK-architecture which allows us to turn complex sampling tasks with sign problems into real ones or complex ones with a mildly varying phase. We have illustrated the potential of the approach within various single-site problems and the real-time dynamics in the harmonic oscillator. In all these cases the complex PIK-architecture provides the correct solution, also for cases where related approaches such as Complex Langevin  and Lefschetz thimbles fail. Further applications to fermionic quantum field theories, higher dimensional models, and the real-time dynamics of interacting theories are in preparation and will be considered elsewhere. \\[-2ex]

\textit{Acknowledgements.---} We thank Gert Aarts, Kirill Boguslavski, Lucas Engel, Kenji Fukushima, Thore K. Joswig, Konrad Kockler, Timoteo Lee, Michael Mandl, Johannes V. Roth, Alexander Rothkopf, Manfred Salmhofer, Sören Schlichting, Christian Schmidt, Erhard Seiler, Denes Sexty, Ion-Olimpiu Stamatescu and Lingxiao Wang for discussions. This work is funded by the Deutsche Forschungsgemeinschaft (DFG, German Research Foundation) under Germany’s Excellence Strategy EXC 2181/1 - 390900948 (the Heidelberg STRUCTURES Excellence Cluster), the Collaborative Research Centre SFB 1225 - 273811115 (ISOQUANT) and the Collaborative Research Centre SFB 1238 - 277146847 (Control and Dynamics of Quantum Materials, project C02). This work was completed during a stay of the three authors at the University of Tokyo, supported by the KAKENHI grant "Foundation of Machine Learning Physics."


\bibliographystyle{apsrev4-2}
\bibliography{refs}

\newpage

\renewcommand{\thesubsection}{{S.\arabic{subsection}}}
\setcounter{section}{0}
\titleformat*{\section}{\centering \Large \bfseries}

\onecolumngrid


\section*{Supplemental Materials}

The supplemental materials provide the derivation of our theoretical set-up, further examples and details of the numerical implementation. Additional material on the set-up can be found in the works on the PIRG approach \cite{Ihssen:2024ihp} and the PIK work on sampling with real distributions \cite{Ihssen:2025ybn}. 

In \Cref{app:ComplexPIK} we detail the derivation of the complex PIK-architecture and provide further explanations. \Cref{app:ZeroDimExamples} contains additional material for the single-site model in the main text as well as several more examples of zero-dimensional field theories which illustrate different features of complex PIKs. Specifically the full analytic access allows us to discuss basic features of the deformation of the integration manifolds. In \Cref{app:FreeTheoryAnyd} we provide computational details for the real-time applications of complex PIKs in quantum mechanics and higher dimensions. Finally, \Cref{app:PIKNumerics} describes the numerical implementation of the complex PIK-architecture.

\subsection{Complex PIK-architecture}
\label{app:ComplexPIK}

In this supplement we provide details on the complex PIK-architecture \labelcref{eq:ComplexPIKwoSignProblem}, and in particular  derivations and further discussions that would have sidetracked from the line of arguments in the main text. The complex PIK-architecture uses a flowing distribution $p_t(\phi)$ with 
\begin{align}
p_t(\phi) = e^{-S_t(\phi)}\,,\qquad \qquad S_t(\phi)= \hat S_t(\phi)+ \log {\cal N}_t\,, \qquad \textrm{with}\qquad \frac{\partial {\cal N}_t }{\partial\phi}=0\,.
	\label{eq:Distribution+LikelihoodApp}
\end{align}
The flow equation for the likelihood $-S_t$ is provided in \labelcref{eq:WF-dotphi} and we recall it here for the sake of completeness, 
\begin{align}
\dot{\phi}_{t,i} (\phi) \,\frac{S_t(\phi)}{\partial \phi_i}- \frac{\partial  \dot{\phi}_{t,i}(\phi)}{\partial \phi_i}=-\frac{d\,S_t(\phi)}{dt} \,. 
\label{eq:WF-dotphiApp}
\end{align}
In comparison to \labelcref{eq:WF-dotphi} we have reshuffled the terms and also display the dependence on $i\in{\cal D}$. The form in \labelcref{eq:WF-dotphiApp} makes it more apparent that the flow defines a differential map $\dot\phi(\phi;S_t)$ for $\phi\in{\cal M}_t$ and hence for ${\cal M}_t$ itself, 
\begin{align} 
	{\cal M}_t \stackrel{\dot \phi}{\longrightarrow } {\cal M}_{t+\Delta t}\,,\qquad {\cal M}_{t+\Delta t} = {\cal M}_t\cup \Delta {\cal M}_{t,\Delta t}\,,\qquad \textrm{with}\qquad
	\Delta {\cal M}_{t,\Delta t}=\int_t^{t+\Delta t}  ds \, \partial_s {\cal M}_s
	\,. 
\label{eq:DiffMapApp}
\end{align} 
For complex ${d}S_t(\phi)/dt $ and/or $\partial S_t(\phi)/\partial \phi$ it follows readily that $\dot\phi(\phi)$ is complex and so is $\Delta{\cal M}_{t,\Delta t}$: 
${\cal M}_{t+\Delta t}$ is the union of the oriented manifolds ${\cal M}_{t}$ and $\Delta {\cal M}_{t,\Delta t}$ in the complexification ${\cal C}$ of the underlying manifold ${\cal M}$, that is ${\cal M}_t, {\cal M}_{t+\Delta t}, \Delta {\cal M}_{t,\Delta t}\subset {\cal C}$. In our lattice example \labelcref{eq:ExampleTheory}, this simply entails ${\cal M}_t\,,\, {\cal M}_{t+\Delta t}\,,\, \Delta {\cal M}_{t,\Delta t}\subset \mathbbm{C}^{\cal D} $. 

\begin{figure}[t]
	\centering
	\begin{subfigure}{.35\linewidth}
		\centering
		\includegraphics[width=\linewidth]{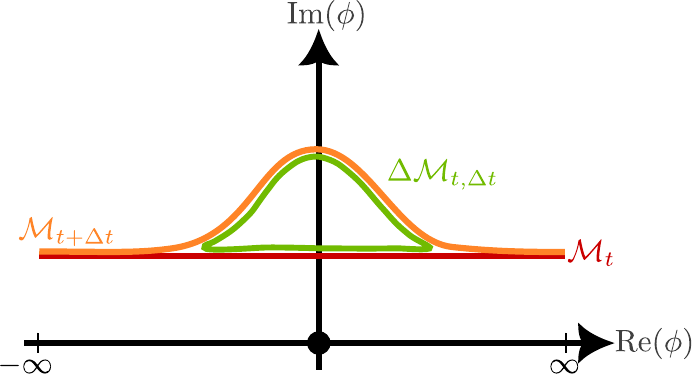}
		\subcaption{Oriented closed hyper-surface $\Delta{\cal M}_{t,\Delta t}$.}
	\end{subfigure}%
	\hspace{3cm}%
	\begin{subfigure}{.35\linewidth}
		\centering
		\includegraphics[width=\linewidth]{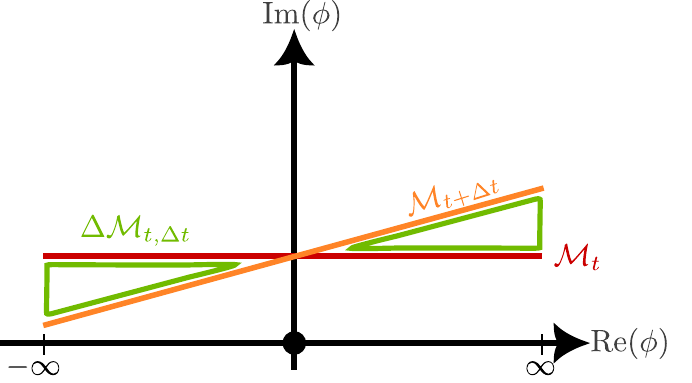}
		\subcaption{Oriented open hyper-surface $\Delta{\cal M}_{t,\Delta t}$.}
	\end{subfigure}%
	\caption{Illustration of different scenarios for the change in the manifolds ${\cal M}_t$ under the application of the PIK ${\dot \phi}_t$.\hspace*{\fill}}
	\label{fig:ClosedDifference}
\end{figure}

As we consider \textit{smooth} maps $\dot\phi(\phi)$, the manifolds $\Delta {\cal M}_{t,\Delta t}$ have no boundaries. For maps $\dot\phi$, that decay sufficiently fast for large amplitudes of $\phi$, the 
manifolds $\Delta{\cal M}_{t,\Delta t}$ are oriented \textit{closed} hyper-surfaces. We also consider \textit{open} oriented hyper-surfaces as generically $\dot \phi(\phi)$ 
exhibits a term that is linear in $\phi$. Note that the latter also accommodates global rotations of ${\cal M}_t$ in the complex manifold ${\cal C}$. In our lattice example this is provided by $\phi\to e^{i \theta_t}\,\phi $, which is discussed further in \Cref{app:RealTimeProxy0d}. Both cases are illustrated for ${\cal M}={\cal M}_0=\mathbbm{R}$ (that is $\varphi,\phi_0\in \mathbbm{R})$ and ${\cal M}_t\subset \mathbbm{C}$ in \Cref{fig:ClosedDifference}. Note that generically ${\cal M}_1 \neq \mathbbm{R}$ and hence $\phi_1\notin \mathbbm{R}$.

With these preparations we formulate a constraint for the differential map $\dot\phi(\phi)$ of the manifold ${\cal M}_t$: it should not sweep over singular points or domains of the distribution $p_t(\phi)$ and the set of considered observables ${\cal O}(\phi)$, 
\begin{align} 
	\oint\limits_{\Delta {\cal M}_{t,\Delta t}}  d\mu_t\,p_t(\phi)\,{\cal O}(\phi)=0 \,, 
	\label{eq:InvertiblezeroApp}
\end{align}
with the closed or open hyper-surface $\Delta {\cal M}_{t,\Delta t}$ defined in \labelcref{eq:DiffMapApp}. We have used $d\mu_t$ also for the measure $d\mu({\cal M}_{t,\Delta t})$ of the hyper-surface in a slight abuse of notation. 

\Cref{eq:InvertiblezeroApp} follows in the absence of singularities of $p_t(\phi)\,{\cal O}(\phi)$ inside $\Delta {\cal M}_{t,\Delta t}$. A particular rôle is played by singularities at $|\phi|\to\infty$. Then, $\Delta{\cal M}_{t,\Delta t}$ are open hyper-surfaces and \labelcref{eq:InvertiblezeroApp} implies
\begin{align}
	\lim_{|\phi| \to \infty} \Re[S_t(\phi)]>0 \qquad \textrm{for} \qquad \phi\in {\cal M}_t\,.
	\label{eq:CloseContourApp}
\end{align}
We emphasise that the integrand in \labelcref{eq:InvertiblezeroApp} is analytically accessible and we discard all PIRG pairs $(S_t,\dot\phi)$ that violate \labelcref{eq:InvertiblezeroApp}. In summary,   
\labelcref{eq:WF-dotphiApp,eq:DiffMapApp} with \labelcref{eq:CloseContourApp} comprise a general differential reparametrisation of the path integral with a smooth deformation of the integration manifold ${\cal M}_t$ if $\dot\phi_t(\phi)$ is non-singular (holomorphic) and invertible on an infinitesimal neighbourhood of ${\cal M}_t$.

%

\subsubsection{Derivation of the complex PIK-architecture} 
\label{app:DerivationComplexPIK}

The PIK-architecture in \cite{Ihssen:2025ybn} maps a sampling problem with the distribution $p(\varphi)$ to an efficient sampling with a simple distribution $p_0(\phi)$ in a differential way via \labelcref{eq:WF-dotphiApp} with distributions $p_t(\phi)$. Specifically, we can define 
operators ${\cal O}(\varphi_t(\phi))$, whose expectation values with the distribution $p_t$ are $t$-independent and provide the expectation value of ${\cal O}(\phi)$ in the underlying distribution $p_1(\phi)=p(\phi)$ for $t=1$. These operators are given by 
\begin{subequations} 
\label{eq:ComplexPIKwithOps}
\begin{align} 
	\left\langle {\cal O}\bigl(\varphi_t(\phi)\bigr)\right\rangle_{p_t} = \int\limits_{{\cal M}_t}   d\phi  \, p_t(\phi) \,{\cal O}(\varphi_t(\phi))\,, 
	\label{eq:ExpOt}
\end{align} 
with 
\begin{align} 
	\frac{d}{d t} \left\langle {\cal O}\bigl(\varphi_t(\phi)\bigr)\right\rangle_{p_t} =0\,,\qquad \textrm{and}\qquad \left\langle {\cal O}\bigl(\varphi^{\ }_1(\phi)\bigr)\right\rangle_{p_1} =
	\int\limits_{{\cal M}_1}   d\phi  \, p(\phi) \,{\cal O}(\phi)\,.
	\label{eq:dtExpOt0}
\end{align} 
\Cref{eq:dtExpOt0} is achieved for the non-trivial $t$-dependent field $\varphi_t(\phi)$ with the properties   
\begin{align} 
	\frac{ d \varphi^{\ }_t(\phi)}{d t} = - \dot \phi_t(\phi)\frac{\partial \varphi^{\ }_t(\phi)}{\partial\phi}\qquad \textrm{and}\qquad \varphi^{\ }_1(\phi)=\phi\,.	
	\label{eq:IntegrateMapApp}
\end{align}
\end{subequations} 
The field $\varphi_t(\phi)$ can be obtained by integrating the differential map from $t=1$ to $t$ with the initial condition $\varphi_1(\phi)=\phi$ in \labelcref{eq:IntegrateMapApp}, 
\begin{align} 
	\varphi^{\ }_t(\phi)= \phi -  \int_1^t d s\, \dot \phi_{s}(\phi)\frac{\partial \varphi^{\ }_{s}(\phi)}{\partial\phi} \,.
	\label{eq:varphitApp}
\end{align}
\Cref{eq:varphitApp} assumes the global existence of the differential map \labelcref{eq:IntegrateMapApp} for all $t$. We emphasise that the existence of \labelcref{eq:varphitApp} for $t\in [0,1]$ is readily proven for any PIRG pair  $(S_t(\phi),\dot\phi(\phi))$, as we have to discard all paths with singular solutions. We note in passing that $S_t$ and $\dot\phi$ are not independent and in the present study we mainly consider $\dot\phi(\phi;S_t)$. However, PIRG pairs, \cite{Ihssen:2024ihp, Ihssen:2025ybn}, maximise the generality of the approach and the optimisation procedure of finding appropriate PIKfolds \labelcref{eq:PIKfold} 
can make use of this generality. The full scope of PIRG pairs, and the related essential RG \cite{Baldazzi:2021orb, Baldazzi:2021ydj}, is also currently being investigated in the functional RG in the context of complex actions \cite{Ihssen:2022xjv}, optimal expansion schemes \cite{Isaule:2018mxt, Isaule:2019pcm, Ihssen:2023nqd, Bonanno:2025mon}, gauge theories \cite{Ihssen:2025cff} and generalised operator flows \cite{Ihssen:2025hyl}.

The field $\varphi_t(\phi)$ carries the whole map from the theory with the simple distribution $p_0(\phi)$ to that of interest with $p_1(\phi)=p(\phi)$. Trivially, the second relation in \labelcref{eq:IntegrateMapApp} arranges for the second property in \labelcref{eq:dtExpOt0}: the expectation value \labelcref{eq:ExpOt} reduces to that of ${\cal O}(\varphi)$ with the distribution $p(\varphi)$ sampled on the manifold ${\cal M}_1$ for $t=1$. With \labelcref{eq:EquivalenceM} this leads us to 
\begin{align} 
\int\limits_{{\cal M}}   d\phi  \, p(\phi) \,{\cal O}(\phi)=\left\langle {\cal O}\bigl(\varphi^{\ }_1(\phi)\bigr)\right\rangle_{p_1} \,. 
\label{eq:TrueSampling}
\end{align}
\Cref{eq:TrueSampling} entails that at $t=1$ the original sampling problem is solved. The first relation in \labelcref{eq:IntegrateMapApp} ensures the first relation in \labelcref{eq:dtExpOt0}: the $t$-independence of the expectation value of ${\cal O}(\varphi_t(\phi))$. This can be shown by considering the $t$-derivative of ${\cal O}(\varphi_t(\phi) )$, 
\begin{align}
	\frac{d}{d t} 	{\cal O}\bigl(\varphi_t(\phi)\bigr)=- \dot \phi_t(\phi) \frac{\partial{\cal O}\bigl(\varphi_t(\phi)\bigr)}{\partial \phi}\,. 
	\label{eq:dOdt}	
\end{align}
Using \labelcref{eq:dOdt} for the $t$-derivative in \labelcref{eq:ExpOt} leads us to 
\begin{align} 
	\frac{d}{d t}	\int\limits_{{\cal M}_t}   d\phi  \, p_t(\phi) \,{\cal O}(\varphi_t(\phi))= \int\limits_{{\cal M}_t}   d\phi  \, \frac{d}{d t} \Bigl[ p_t(\phi)	 \,{\cal O}(\varphi_t(\phi))\Bigr]= - \int\limits_{{\cal M}_t}   d\phi  \, \frac{\partial}{\partial \phi } \Bigl[ \dot \phi\, p_t(\phi)	 \,{\cal O}(\varphi_t(\phi))\Bigr] =0\,.
	\label{eq:DiffEquivalenceSamplingApp}
\end{align} 
For the first equality in \labelcref{eq:DiffEquivalenceSamplingApp} we have used \labelcref{eq:InvertiblezeroApp}. The second equality follows with \labelcref{eq:dOdt} and the flow \labelcref{eq:WF-dotphiApp}, or rather its form \labelcref{eq:WegnerEquation} for ${d}p_t/dt$. The final equality follows from the vanishing of $p_t(\phi) \,{\cal O}(\varphi_t(\phi))$ on the boundary $\partial {\cal M}_t$ of the manifold ${\cal M}_t$. 

\Cref{eq:ComplexPIKwithOps,eq:TrueSampling} can be used to formulate the PIK-architecture concisely in terms of expectation values of general operators ${\cal O}(\varphi)$, 
\begin{subequations} 
	\label{eq:ComplexPIKwoSignProblemApp}
\begin{align} 
		\int\limits_{{\cal M}}  {\cal D}\varphi \, p(\varphi) \,{\cal O}(\varphi)= \int\limits_{{\cal M}_0}  d\phi  \,p_0(\phi)\,{\cal O}\bigl(\varphi(\phi)\bigr)\,, 
		\label{eq:EquivalenceSamplingApp}
\end{align} 
with $\varphi(\phi)=\varphi^{\ }_0(\phi)$ in \labelcref{eq:varphitApp}, 
\begin{align} 
	\varphi(\phi)= \phi + \int_0^1 ds \, \dot \phi_{s}(\phi)\frac{\partial \varphi^{\ }_{s}(\phi)}{\partial\phi}\,, 
	\label{eq:phi0App} 
\end{align}
and the map $\varphi(\phi)$ has to be holomorphic in the PIKfold ${\cal P}$ \labelcref{eq:PIKfold}. 
\end{subequations} 
Note that \labelcref{eq:ComplexPIKwoSignProblemApp} can readily be extended to meromorphic maps $\varphi(\phi)$, which will be discussed elsewhere. 

For the sampling procedure it is more convenient to rewrite \labelcref{eq:phi0App} in terms of the $t$-dependent coordinates $\phi_t$ and 
the change $\dot\phi_t[\phi_t]$: To that end we define $\phi_t$'s that provide a $t$-independent $\varphi_t(\phi_t)$, 
\begin{align} 
\frac{d \varphi^{\ }_t(\phi_{t})}{d t} \stackrel{!}{=} 0\qquad \Rightarrow \qquad \frac{d\phi_{t}}{dt}(\phi_t) = \dot\phi_{t}(\phi_{t})\,,  \qquad \phi_1= \varphi_1(\phi_1)=\varphi_t(\phi_t)\,, 
	\label{eq:Defphit}
\end{align} 
where we have used the $t$-independence of $\varphi_t(\phi_t)$ in the last equality. 
The differential equation for $\phi_t$ is readily integrated. We use the samples $\phi_0$ of the trivial distribution $p_0$ as initial condition and are led to 
\begin{align}
 \phi_{t} =\phi_0 + \int_0^t ds \, \dot \phi_{s} (\phi_{s})\,, 
\label{eq:PhiMapFromCalc}
\end{align}	
with $\phi_s=\phi_s(\phi_0)$ for $s\in[0,1]$. Note that \labelcref{eq:PhiMapFromCalc} allows us to take any $\phi_t$ as the variable but 
for the sampling task with $p_0(\phi)$, $\phi_0$ is the natural choice as it corresponds to samples of the initial distribution. Moreover, we have $\phi_1= \varphi_t(\phi_t)$ for all $t$, \labelcref{eq:Defphit}. Hence, for $t=1$, \labelcref{eq:phi0App} can be rewritten as  
\begin{align}
\varphi(\phi_0) = \phi_0+ \int_0^{1} ds \, \dot \phi_{s} (\phi_{s})\,. 
	\label{eq:varphi-phidotphi}
\end{align}	
This concludes our derivation of the complex PIK-architecture: \Cref{eq:EquivalenceSamplingApp,eq:varphi-phidotphi} are nothing but \labelcref{eq:ComplexPIKwoSignProblem} with $\phi_0=\phi$. The integrated PIK \labelcref{eq:varphi-phidotphi} provides the maps from ${\cal M}_0$ to ${\cal M}_t$ and the set of these manifolds, $\{{\cal M}_t\}$, endowed with the PIK-map, is the PIKfold introduced in \labelcref{eq:PIKfold}. 

We close this supplement with some remarks on the properties and extensions of the PIK-architecture as well as a discussion of the sampling procedure: \\[-2ex] 

The weight-preserving property of the PIK-architecture is discussed in \Cref{app:WeightPIK} and generalisations thereof are introduced in \Cref{app:BeyondPIKfold}. 

We also remark, that the construction readily extends to meromorphic maps \labelcref{eq:varphi-phidotphi}, if the pole contributions are properly taken into account. Note also that $\varphi(\phi)$ being holomorphic for $\phi\in {\cal P}$ does not imply $\phi_t(\phi_0)$ in \labelcref{eq:PhiMapFromCalc} being holomorphic on ${\cal P}$ or even on the sub-PIKfold ${\cal P}_t$ defined in \labelcref{eq:PIKfold}. 

Moreover, the PIK provides a differential map between the manifolds ${\cal M}_t$. We readily conclude that therefore they do not factorise even if the initial manifold ${\cal M}_0$ does. This is because the differential map \labelcref{eq:WF-dotphiApp} does not maintain such a factorisation in general. Moreover, 
we may even start with a non-factorisable manifold ${\cal M}_0$ that can result from an optimisation procedure. In our lattice example \labelcref{eq:ExampleTheory} the lack of factorisation due to $\dot \phi$ is induced by the hopping term. In general it is induced by any correlation of different data points in the data distribution $p_t(\phi)$. Accordingly, the measures $d\mu_t$ do not factorise into separate integrations on each data point. 

Finally, we briefly discuss the sampling procedure. With \labelcref{eq:EquivalenceSamplingApp,eq:varphi-phidotphi}, the sampling task with $p(\varphi)$ on ${\cal M}$ is reduced to an efficient one with $p_0(\phi)$ on ${\cal M}_0$ if $\varphi(\phi)$ is a well-defined map. 
Then, operators can be sampled at any intermediate RG-time $t$ using
\begin{align}
	\langle \mathcal{O}(\phi)\rangle_{p_t} = \frac{1}{N_s} \sum_n \mathcal{O}(\phi_t(\phi_0)) \,,
	\label{eq:SampleOt}
\end{align}
where $\phi_t$ was defined in \labelcref{eq:PhiMapFromCalc} and we sum over $N_s$ samples $\phi_0$ drawn from the initial distribution $ p_0(\phi_0)$. Finally we recall that this sampling task has no sign and overlap problem as discussed around \labelcref{eq:dtWeightPT=0} and in \Cref{app:WeightPIK}.

\subsubsection{Normalisation reloaded} 
\label{app:NormalisationReloaded}

When computing physics-informed kernels $\dot{\phi}_t(\phi)$, the normalisation $\mathcal{N}_t$ of the theory under investigation takes a central rôle. In this appendix, we demonstrate how this rôle is taken up by a linear contribution in the kernel. Moreover, we show how that the computational task computing the normalisation is computationally efficient. To that end, we factor out the normalisation ${\cal N}_t$ in the distribution $p_t$, to wit, 
\begin{subequations} 
\label{eq:hatpNt}
\begin{align}
	\hat p_t(\phi) = e^{-\hat S_t(\phi)} \,, \qquad \textrm{with} \qquad p_t(\phi) = \frac{1}{\mathcal{N}_t} \hat p_t(\phi) \,. 
	\label{eq:DefinitionUnnormalisedPt}
\end{align}
The action $\hat S(\phi)=S(\phi)-S(0)$ was introduced in \labelcref{eq:pvarphi}, and the normalisation ${\cal N}_t$ follows as 
\begin{align}
\mathcal{N}_t = \int_{{\cal M}_t}  d\phi  \, \hat p_t(\phi) \,.
	\label{eq:DefinitionNt}
\end{align}
\end{subequations}
\Cref{eq:DefinitionNt} entails that the normalisation depends on the given action $\hat S(\phi)$, that is ${\cal N}_t = {\cal N}_t[\hat S_t(\phi)]$. Moreover, \labelcref{eq:DefinitionNt} constitutes a hard sampling problem. This sampling problem can be circumvented by reformulating the Wegner equation \labelcref{eq:WegnerEquation} as an equation for the normalisation, see \cite{Ihssen:2025ybn}. To that end we insert the parametrisation \labelcref{eq:DefinitionUnnormalisedPt} of the distribution $p_t(\phi)$ into \labelcref{eq:WegnerEquation}, which leads us to 
\begin{align}
	\left[ \frac{d}{dt} - \frac{d\,\log\mathcal{N}_t}{dt} \right]\, \hat p_t(\phi)= -\frac{\partial}{\partial \phi_i} \left[ \dot \phi_{t,i}(\phi) \, \hat p_t(\phi) \right]\,.
	\label{eq:WegnerAppearanceOfNormalisation}
\end{align}
Solving \labelcref{eq:WegnerAppearanceOfNormalisation} for the flow of ${\cal N}_t$ results in 
\begin{align}
	\frac{d\,\log\mathcal{N}_t}{dt} = -\frac{d\,\hat S_t(\phi)}{dt} + \left[ \frac{\partial}{\partial \phi_{i}} - \frac{\partial \hat S_t(\phi)}{\partial \phi_i} \right]\dot \phi_{t, i}(\phi) \,.
	\label{eq:WegnerEquationConstant}
\end{align}
\Cref{eq:WegnerEquationConstant} implies that the right-hand side is independent of the configuration used. Hence the normalisation can be obtained from evaluating the right-hand side on optimally chosen configurations, picked in view of the solution scheme used for $\dot \phi$. Moreover, the (in)dependence of the right-hand side on the chosen configurations is one of the measures for the systematic error of the solution. 

 We close this supplement with a discussion of the relation between the normalisation and the kernel $\dot{\phi}_t(\phi)$. For the sake of simplicity we restrict ourselves to $\mathbbm{Z}_2$-symmetric theories with fields defined on an unbounded domain, such as the $\phi^4$-theory. To that end we parametrise the kernel as 
\begin{align} 
	\dot\phi_{t,i}(\phi) = \gamma_{\phi \, i \,}{}^{ j}\,\phi_j + \Delta\dot\phi_{t,i}(\phi)\,,\qquad \textrm{with} \qquad \frac{\partial \Delta\dot\phi_{t,i}}{\partial\phi_i}(\phi=0)=0\,,
	\label{eq:Deltadotphi}
\end{align} 
where $\gamma_\phi$ controls the linear contribution and $\Delta\dot\phi_t(\phi)$ contains the remaining terms in $\dot{\phi}_t$. Then, the Wegner equation \labelcref{eq:WegnerEquation} takes the form 
\begin{align}
	\frac{d \,p_t(\phi)}{d t} = -\left(\gamma_\phi{}^i{}_i + \gamma_\phi{}^{ij}\phi_j \frac{\partial}{\partial\phi^i  }\right)\,p_t(\phi)
-\frac{\partial}{\partial\phi_i}\Bigl[ \Delta\dot\phi_{t,i}(\phi) \, p_t(\phi)\Bigr] \,.
	\label{eq:WegnerEquationReparametrised} 
\end{align}
With the parametrisation \labelcref{eq:hatpNt}, this translates into 
\begin{align}
 \frac{d\,\log\mathcal{N}_t}{dt} -\gamma_\phi{}^i{}_i =-	\frac{d\,\hat S_t(\phi) }{dt}  - \left( \gamma_\phi{}^{i j} \phi_j + \Delta \dot{\phi}^i_t (\phi)\right) \,\frac{\partial \hat S_t(\phi)}{\partial \phi^{i}}+ \frac{\partial  \Delta \dot{\phi}^i_t(\phi)}{\partial \phi^i} \,.
	\label{eq:WF-dotDeltaphi}
\end{align}
With \labelcref{eq:Deltadotphi}, $\Delta\dot\phi$ does not contain a linear term in $\phi$ and hence the right-hand side does not contain a field-independent part. This leads us to 
\begin{align} 
\gamma_\phi{}^i{}_i= \frac{d \log {\cal N}_t}{d t} \,, 
	\label{eq:RelgammaphiNt} 
\end{align} 
and 
\begin{align}
	\frac{d\,\hat{S}_t(\phi)}{dt} +\left( \gamma_\phi{}^{i j} \phi_j+\Delta \dot{\phi}^i_t (\phi)\right)\,\frac{\partial \hat S_t(\phi)}{\partial \phi^i}= \frac{\partial  \Delta \dot{\phi}^i_t(\phi)}{\partial \phi^i} \,. 
	\label{eq:WF-Deltadotphi}
\end{align}
This shows that while $\hat S_t$ can be chosen freely, the normalisation ${\cal N}_t$ is determined by $\hat S_t$, that is ${\cal N}_t={\cal N}_t[\hat S_t]$, see \labelcref{eq:DefinitionUnnormalisedPt}. Moreover, this leaves us with a flow of $\hat p_t(\phi)$ without normalisation problem.  

The above properties reflect those of PIRGs for the quantum effective action or rate function, where the constant term in the effective action is uniquely fixed by the demand of global solutions of $\dot\phi_t(\phi)$ at each $t$ \cite{Ihssen:2024ihp}. This constraint translates into a similar one in the current PIK-architecture: the $\hat S$ are constrained to those that admit a $\textrm{global}$ \textit{holomorphic}
 solution of the differential equation \labelcref{eq:WF-dotphi} for the PIK $\dot\phi_t(\phi)$ for $\phi$ in $\phi\in{\cal P}$. While implicit, these constraints are naturally assessed by constructing the solution of the differential map $\dot\phi_t$ for all $t$. A final constraint is given by the existence of the integrated map \labelcref{eq:varphi-phidotphi}.

\subsubsection{Weight-preserving property of the PIKfold and the absence of sign and overlap problems} 
\label{app:WeightPIK}

As discussed in the main text, sign problems originate in oscillations of the complex distribution. Overlap problems are present if one has to draw many samples in the tail of the distribution $p_0(\phi)$ to sample $p(\varphi)$ accurately. In this supplement we show that both problems are absent in the PIK-architecture, which solves both problems by construction. Consequently, the task of sampling without sign and overlap problems in a PIK-architecture is simply that of providing a PIK which leads to a PIKfold \labelcref{eq:PIKfold}: can we integrate 
$\dot\phi$ from $t=0$ to $t=1$, or, alternatively, we can solve the simple ${\cal D}$-dimensional analogue of the differential equation \labelcref{eq:DefofMt}, or similar ones obtained from the complex sampling pair $(p_0\,,{\cal M}_0)$ for all $t\in (0,1]$? 

\Cref{eq:DefofMt,eq:ConstructionMt} follow directly from the weight-preserving property \labelcref{eq:dtWeightPT=0} of the PIK. There, the property has been proven by using the equivalence of expectation values of general operators and using characteristic functions of the flowing subsets ${\cal T}_t$ under investigation.   

Her we consider a differential point of view which is also important for the generalisation of the PIKfold considered in \Cref{app:BeyondPIKfold}. We begin the derivation of the latter by recalling the definition of the 'statistical weight' of a general subset ${\cal I}_t\subset {\cal M}_t$, 
\begin{align} 
	P_t({\cal I}_t)  = \int_{{\cal I}_t}   d\mu_t \, p_t(\phi) \,,\qquad \textrm{with} \qquad P_t({\cal I}_t) \in\mathbbm{C}\,, \qquad \mathrm{and} \qquad P_t({\cal M}_t)=1\,.
	\label{eq:WeightPS}
\end{align}
In general, these weights are complex and the weight of the total data set is normalised to unity by construction of the PIK. The flow of the weight $P_t({\cal I}_t) $ has two contributions, one from the flow of the distribution $p_t$ and one from the flow of the boundary of the subset ${\cal I}_t$. The flow of the submanifold ${\cal I}_t$ has again two contributions: One comes from the direct $t$-variation of the boundary of ${\cal I}_t$ in ${\cal M}_t$ which is at our disposal. The other comes from the flow of the manifold ${\cal M}_t$ in the complexification ${\cal C}$ of the original manifold $\cal M$ (in our lattice field theory example $\mathbbm{C}^{\cal D}$). However, the flow of the manifold is a closed or open hyper-surface, see \Cref{fig:ClosedDifference} and the respective integral vanishes if $p_t(\phi)$ has no singularity inside the hyper-surface. This leads us to   
\begin{align} 
	\frac{d\,P_t({\cal I}_t)}{dt}  = \int_{\partial_t {\cal I}_t}   d\mu_t \, p_t(\phi)-   \int_{{\cal I}_t}   d\mu_t \, \frac{\partial}{\partial \phi_i} \left[\dot\phi_{t,i} p_t(\phi)\right] = \int_{\partial {\cal I}_t}  d{\cal {\cal S}}_i(\phi)\, \left[\sigma_{t,i}(\phi)-\dot\phi_{t,i}(\phi)\, \right]\, p_t(\phi)\,, 
	\label{eq:dtWeightPS}
\end{align}
where $\sigma_{t,i}(\phi)$ corresponds to rate of the $t$-variation of the manifold ${\cal I}_t$. It has two contributions: one from the flow of the manifold ${\cal M}_t$, which is simply $\dot \phi$, and the other is an explicit one that is up to our disposal,  
\begin{align} 
	\sigma_{t,i}(\phi)= \dot \phi_{t,i} (\phi)+\Delta 	\sigma_{t,i}(\phi)\,.
	\label{eq:FlowT}
\end{align}
Inserting \labelcref{eq:FlowT} into \labelcref{eq:dtWeightPS} leads us to 
\begin{align} 
	\frac{d\,P_t({\cal I}_t)}{dt}  = \int_{\partial {\cal I}_t}  d{\cal {\cal S}}_i(\phi)\, \Delta \sigma_{t,i}(\phi)\, p_t(\phi)\,. 
	\label{eq:dtWeightPSDeltamu}
\end{align}
\Cref{eq:dtWeightPSDeltamu} entails that the weight of general sub-manifolds ${\cal I}_t$ only changes if the boundaries of the manifold are changed by hand. Hence, if we do not deform the manifold explicitly, i.e.~$\Delta \sigma_{t,i}(\phi)\equiv 0$, we obtain flowing sub-manifolds ${\cal T}_t$ which are images of some original ${\cal T}_0$ with the integrated map of the PIK $\dot\phi$. For such manifolds ${\cal T}_t$, \labelcref{eq:dtWeightPSDeltamu} reduces to 
\begin{align} 
	\frac{d\,P_t({\cal T}_t)}{dt}  = 0\,,\qquad P_t({\cal T}_t)  = \int_{{\cal T}_t}   d\mu_t \, p_t(\phi) \,.
	\label{eq:dtWeightPT=0Supplement}
\end{align}
\Cref{eq:dtWeightPT=0Supplement} shows the weight-preserving property of the PIK. Consequently, there is no overlap problem. Moreover, if the sampling pair $(p_0\,,{\cal M}_0)$ has no sign problem, all pairs $(p_t\,,{\cal M}_t)$ are sign-problem free. However, the defining differential equation for ${\cal M}_t$, which follows from the weight-preserving property, may have no solution. Again we resort to the one-dimensional example with a real sampling pair $(p_0\,,{\cal M}_0)$ that leads to \labelcref{eq:DefofMt}, which we recall here for the sake of convenience,  
\begin{align} 
	\frac{d\phi_I}{d \phi_R} =-\frac{\textrm{Im}\left[p_t(\phi)\right]}{\textrm{Re}\left[p_t(\phi)\right]} \,.
	\label{eq:DefofMtSupplement}
\end{align} 	
For $\textrm{Re}[p_t(\phi)]=0$, the right-hand side hits a singularity and no smooth solution for the differential equation can be found. 
Note that $\textrm{Re}[p_t(\phi)]=0$ on a sampling manifold is one of the signatures of a sign problem. As mentioned in the main text, we did not encounter it for the examples considered here, but it can be shown that it is rather generic for fermionic theories. In these cases the sampling pair $(p_0\,,{\cal M}_0)$ has to be changed, which will be considered in \cite{PIKF2026}.

\subsubsection{Sampling beyond the PIKfold} 
\label{app:BeyondPIKfold}

An important generalisation is provided by augmenting the flow of the sampling pair $(p_t\,,{\cal M}_t)$ with 
an additional deformation of the sampling manifold that is not triggered by the PIK $\dot\phi$. This leads us to sampling pairs 
$(p_t, {\cal G}_t)$ with 
\begin{align} 
	{\cal G}_t={\cal M}_t\cup\Delta {\cal G}_t\,,
\label{eq:ManifoldG}
\end{align}
with $\Delta {\cal G}_t$ being the closed or open hyper-surface between the slice ${\cal M}_t$ of the PIKfold and ${\cal G}_t$. This is similar to \labelcref{eq:DiffMapApp}, the difference being that $\Delta{\cal G}_t$ does not originate in the flow of the PIKfold but is an explicit flow of the manifold which is up to our disposal. If $p_t(\phi)\,{\cal O}(\phi)$ has no singular points in $\Delta {\cal G}_t$, we conclude 
\begin{align} 
	\oint\limits_{\Delta {\cal G}_{t}}  d\mu_t \,p_t(\phi)\,{\cal O}(\phi)=0 \,. 
	\label{eq:IntDeltaG0}
\end{align}
In a slight abuse of notation we have used $d\mu_t$ for the measure $d\mu(\Delta {\cal G}_{t})$ of the hyper-surface $\Delta {\cal G}_{t}$. 
\Cref{eq:IntDeltaG0} entails that sampling with $\{p_t\}$ can either be done on the PIKfold \labelcref{eq:PIKfold} or on the generalised set of sampling manifolds $\{{\cal G}_t\}$, 
\begin{align} 
	 \int\limits_{{\cal M}}   d\mu  \, p(\varphi) \,{\cal O}(\varphi)= \int\limits_{{\cal G}_t}   d\mu_t  \, p_t(\phi) \,{\cal O}(\phi_t(\phi))\,, 
	\label{eq:O=O}
\end{align} 
for all $t$. Here, $\phi_t(\phi)$ is given by \labelcref{eq:PhiMapFromCalc} and \labelcref{eq:O=O} holds true if $\phi_t(\phi)$ has no singular points in $\Delta{\cal G}_t$. 

However, there is a price to pay for this additional freedom: the loss of the weight-preserving property. This is readily seen by applying \labelcref{eq:dtWeightPS} to sub-manifolds ${\cal I}_t\subset {\cal G}_t$. In this case, $\Delta \mu_t$ in \labelcref{eq:FlowT} simply is the characteristic function of the change of $\Delta {\cal G}_t$ and we find 
\begin{align} 
	P_t({\cal I}_t)= P_0({\cal I}_0) + \int\limits_{\Delta{\cal G}_t \cap {\cal I}_t} \hspace{-.3cm} d\mu_t\,  p_t(\phi)\,. 
	\label{eq:dtWeightPS-IsubG}
\end{align}
In conclusion, deforming the PIKfold with the flow comes at the price of loosing the weight-preserving property in the difference hyper-surface $\Delta {\cal G}_t$. Importantly, the weight of intervals ${\cal I}_t$ with ${\cal I}_t\cup {\cal T}_t\subset \Delta{\cal G}_t$ is preserved. Here, ${\cal T}_t$ is the image of ${\cal I}_0$ with the integrated flow. This entails in particular that the weight of such intervals is real, if the underlying PIKfold is real. This can be used for removing residual sign problems as occur in the Lefschetz thimble approach. In such a case, $\Delta {\cal G}_t$ is the union of hyper-manifolds that envelop these singular points that are the mirrors of regimes that carry relative phases for the original sampling $(p(\varphi)\,,\,{\cal M})$, 
\begin{align} 
	\Delta {\cal G}_t=\sum_i \Delta {\cal G}_{t,i}\,. 
\label{eq:DeltaGPhase}
\end{align}
Importantly, the weight of the single hyper-surfaces $\Delta {\cal G}_{t,i}$ vanishes and the weight of ${\cal M}_t\cap \Delta {\cal G}_{t,i}$ is real for a real PIKfold. Importantly, such a procedure potentially removes a residual sign problem and we call such a set of manifolds a \textit{generalised} PIKfold $g{\cal P}$, 
\begin{align} 
	g{\cal P}=g{\cal P}_1\,,\qquad g{\cal P}_t=\{{\cal G}_s\,|\,s\in[0,t]\}\,,
	\label{eq:PIKfold2}
\end{align} 
endowed with the holomorphic map $\varphi(\phi)$ on $g{\cal P}$. 

We close this supplement with the remark that these deformations can be used to remove the singular points in strong PIKfolds that occur as singularities of \labelcref{eq:DefofMtSupplement} and variants thereof. Alternatively, as mentioned at the end of \Cref{app:DerivationComplexPIK}, one may extend the PIKfold to meromorphic functions \labelcref{eq:varphi-phidotphi} and add the respective pole contributions. Both procedures lead us to generalised PIKfolds $g{\cal P}$. The treatment of the singular points will be considered in \cite{PIKF2026}.

\subsection{The complex PIK-architecture in the landscape of approaches to the sign-problem} 
\label{app:FFR}

Naturally, the PIK-architecture has many conceptual and computational connections to other approaches to the sampling of complex distributions with sign problems. A detailed comparison will be provided elsewhere, here we only briefly discuss some important connections that also allow us to emphasise the differences. Before we embark on this journey we would like to emphasise that this discussion is aimed at readers with some or even detailed familiarity of these approaches. We cannot augment this comparison with an introduction to all these approaches in this supplement, and hence the discussion below assumes a more detailed understanding of the other approaches. A full comparison, concentrating more on the potential exploitation of the embedding of the other approaches in the PIK-architecture, will be provided elsewhere.

\subsubsection{Complex Langevin equations, diffusion models and the complex PIK} 
\label{app:CLE-PIK}

For a general overview of the CLE, see e.g.~the reviews~\cite{Seiler:2017wvd, Aarts:2017hqp, Berger:2019odf}. To begin our comparative discussion, note that the defining equation of the PIK approach is the flow equation \labelcref{eq:WegnerEquation} for the flowing distribution $p_t(\phi)$, that can be understood as a generalised diffusion equation for $p_t(\phi)$. Indeed, if we allow for an operator kernel $\dot\phi$, 
\begin{align}
	\dot\phi_{\textrm{op}}(\phi)= \dot\phi(\phi) + K(\phi) \frac{\partial}{\partial \phi }\,,
	\label{eq:Opdotphi}
\end{align}
the equation accommodates kernel CLE, see e.g.~\cite{Boguslavski:2022dee,Alvestad:2023jgl}: For $\dot\phi= \partial S/\partial \phi$ and $K_{ij}(\phi)=\delta_{ij}$, the flow equation \labelcref{eq:WegnerEquation} reduces to the standard Fokker-Planck equation underlying complex stochastic quantisation. The operator part in \labelcref{eq:Opdotphi} induces shifts in field space with a weight $K(\phi)$ and, hence, in the general case, it induces a field-dependent coloured noise in the derived stochastic Langevin equation for configurations $\phi$. 

Diffusion models naturally arise from a combination of kernel CLE and a more general $\dot\phi$. However, instead of computing the respective PIK $\dot\phi_\textrm{op}$, or rather its coefficients $(\dot\phi\,,\,K(\phi))$, both are learned. This procedure is subject to the two out-of-domain problems of generative machine learning architectures discussed at length in \cite{Ihssen:2025ybn}. 

In both, kernel CLE and diffusion models, the kernel is used for controlling the  'detours' in the complex plane triggered by the stochasticity. From the standpoint of the PIK approach one may interpret these detours as deviation from the weight-preserving property of the PIKfold and the kernel optimisation is bound to minimise these detours. Moreover, in our opinion the embedding of kernel CLE or diffusion models in the PIK-architecture provides a conceptual backbone for this optimisation procedure. Finally, we add that the absence of the noise term eliminates the danger of wrong fixed point solutions which makes the assessment of CLE fixed point so intricate (boundary terms). In summary, it is suggestive that the analytic PIK-access to these maps can be used to augment the CLE approach as well as diffusion models with crucial optimisation steps.

\subsubsection{Lefschetz thimbles and the complex PIK} 
\label{app:LT-PIK}

The deformation of the sampling manifold induced by the PIK is a natural link to the Lefschetz thimble approach, for reviews see e.g.~\cite{Scorzato:2015qts,Berger:2019odf,Alexandru:2020wrj}, for the related resurgence approach see e.g.~\cite{Dunne:2025mye}. While the PIKfold is defined by \labelcref{eq:WF-dotphi,eq:WF-dotphiApp}, the reformulation of \labelcref{eq:WegnerEquation} as a differential equation for the PIK $\dot\phi$, the respective deformation in Lefschetz thimbles evolves the original sampling manifold ${\cal M}$ with $\partial \bar S/ \partial \bar \phi$, where $\bar\phi$ is the complex conjugate of $\phi$ in the complexification ${\cal C}$ of ${\cal M}$.  This deformation flows the sampling manifold towards a in general \textit{disjunct} union of manifolds where the action admits a constant phase. In case the complexification ${\cal C}$ of the original sampling manifold admits several contributing thimbles with different phases, all these contributions have to be summed over. This leads to the \textit{residual} or \textit{global} sign problem. In contradistinction, the PIKfold is weight-preserving. For real PIKfolds with a constant Jacobian of the PIK-map, the two approaches are equivalent if the theory only admits one contributing thimble. This is for example the case in \Cref{fig:NullStar}, where the contour is purely a rotation of the original sampling manifold and $S_t$ is chosen such that $\dot\phi$ only induces this rotation. If several contributing thimbles are present, such as in the example with the action \labelcref{eq:Action0d}, the real PIKfold has no relative phase. In the language of the thimble approach this is similar to pulling back all thimble contributions to a main thimble. In contradistinction to the LT approach this map is provided analytically by construction and the resulting sampling manifold for the example action  \labelcref{eq:Action0d} is depicted in 	\Cref{fig:Showcase} together with the Lefschetz thimbles. It is suggestive that the analytic PIK-access to these maps can be used to augment the LT approach specifically in terms of resolving the residual sign problem.

\subsubsection{Dual variables and the complex PIK} 
\label{app:DV-PIK}

The relation of the complex PIK-architecture to the approach via dual variables, see e.g.~the reviews~\cite{Gattringer:2016kco, Berger:2019odf}, is provided by the PIK-property, that it induces a variable transformation in field space by construction. Indeed, this is one of the first and most successful applications of the generalised flow equations \cite{Wegner_1974, Pawlowski:2005xe}. Specifically, they have been introduced in \cite{Gies:2001nw, Gies:2002hq} for bilinear composite fields to accommodate resonant bilinear bound states in the flow of the effective action $\Gamma_t$. Then, instead of computing $\dot\phi$ from $S_t$, one is solving $\Gamma_t$ for a given  $\dot\phi$. General composites along this lines have been considered in \cite{Lamprecht2007, Isaule:2018mxt, Isaule:2019pcm, Fukushima:2021ctq}, as well as in the essential RG \cite{Baldazzi:2021ydj, Baldazzi:2021orb}. 

The PIRG approach \cite{Ihssen:2023nqd, Ihssen:2024ihp, Ihssen:2025cff, Ihssen:2025hyl} has generalised this approach with a fundamental change of perspective which also underlies the PIK-architectures: instead of solving the flows for a generating function such as the distribution $p_t(\phi)$ or its likelihood $S_t$, or the effective action/rate function $\Gamma_t$, we solve the flow for the PIRG pair $(F_t, \dot \phi_t)$ with $F_t=S_t,\Gamma_t,....$. In general, this task can be distributed among both $F_t$ and $\dot\phi_t$. In \cite{Ihssen:2025ybn} and here we have used a given $F_t=S_t$ and solved only for $\dot\phi$. In turn, in \cite{Ihssen:2024ihp} we have discussed examples the task of computing $\Gamma_t$ was only partially redistributed to solving for $\dot\phi_t$. Hence, using such a procedure in the PIK-architecture allows us to parametrise the pair $(S_t\,,\,\dot\phi_t)$ such that a systematic search for optimised dual variables can be implemented. It is suggestive that the analytic PIK-access to these maps and the weight-preserving property if the PIKfold can be used to augment the approach with dual variables for finding optimal ones or any at all.

\subsubsection{Concluding remarks}
\label{app:PIKrules}

In the above discussion of different approaches to theories with complex distributions we have indicated how these approaches can be embedded into the complex PIK-architecture. While this offers the possibility to augment these approaches with the structural insight and computational simplifications inherent to the PIK approach, it also should be emphasised that one can utilise the computational advances made in these approaches within the PIK approach. We hope to report on these two aspects in the near future.

\subsection{Solving sign problems for benchmark systems in $d=0$ dimensions}
\label{app:ZeroDimExamples}

In this supplement we provide a comprehensive overview of the computations within general $\phi^4$-theories in $d=0$ dimensions. In \Cref{app:FreeTheory0d} we consider a zero-dimensional free theory with a complex current. This setup allows us to solve the Wegner equation analytically and to exemplify some of the properties of PIKs discussed around \labelcref{eq:Idenity}. In \Cref{app:ComplexZ2} and \Cref{app:ComplexZ2+lin} we extend this discussion to systems which are only numerically accessible, i.e.~to general $\phi^4$-theories with complex couplings. The former section allows for a comparison with analytically known thimble structures in $\mathbbm{Z}_2$-symmetric theories, whereas in the latter we augment the $\phi^4$-theory with a linear term, whose coupling is adjusted such that it maximises the sign problem. Again, we find that the simple linear PIK-architecture suffices to perform the sampling task. Finally, we discuss the zero-dimensional proxy of a real-time $\phi^4$-theory in \Cref{app:RealTimeProxy0d}. For this example, the physical distribution is a pure phase, which is reflected in PIKs $\dot \phi$ that do not decay for $\phi \to \infty$.

\subsubsection{Free theory: Adding a complex current in zero dimensions}
\label{app:FreeTheory0d}

As a first example of an analytically tractable, sign-problem free PIKfold, we consider a zero-dimensional free theory with a real mass $m^2 \in \mathbbm{R}_+$ and add a complex current $i t \phi$. The action reads
\begin{align}
	\hat S_t (\phi) = \frac{1}{2} m^2 \phi^2 + i t \phi  \,.
	\label{eq:ExFree_1}
\end{align}
The trajectory starts with a simple Gaussian at $t=0$ and successively adds a purely imaginary current to the action as $t > 0$. The defining equation for the PIK follows by inserting \labelcref{eq:ExFree_1} into the Wegner equation \labelcref{eq:WF-dotphi} and reads
\begin{align}
	\frac{\partial \dot \phi_{t}(\phi)}{\partial \phi} - \dot \phi_{t}(\phi)\; (m^2 \phi+ i t) = i \phi + \frac{d\,\log \mathcal{N}_t}{dt} \,.
\end{align}
This differential equation has an analytic solution, which can be obtained from the method of variation of parameters
\begin{align}
	\dot \phi_t (\phi) = & -\frac{i}{m^2}  +	e^{\frac{m^2}{2}\left(\phi + i \frac{t}{m^2}\right)^2} \left[c_0 + \sqrt{\frac{\pi}{2 m^2}}\left(\frac{t}{m^2}+\partial_t \log \mathcal{N}_t \right) \erf\left(\frac{m^2 \phi + it}{\sqrt{2 m^2}}\right)\right]
	\,.
	\label{eq:GeneralSolutionFree1}
\end{align}
It remains to determine two constants: the integration constant $c_0$, as well as the change of normalisation $\mathcal{N}_t$. Both are fixed by requiring the finiteness of $\dot\phi_t$ for all values of $\phi$, in particular as $|\phi| \to \infty$. Hence we obtain $c_0 = 0$ and
\begin{align}
	\frac{d \log \mathcal{N}_t}{dt}  = -\frac{t}{m^2} \qquad  \Rightarrow \qquad \log \mathcal{N}_t = - \frac{t^2}{2 m^2} + \log \mathcal{N}_0\,,
\end{align}
where $\mathcal{N}_0 = \sqrt{\frac{2 \pi}{m^2}}$, which corresponds to the normalisation of the original Gaussian. The exponential prefactors are a characteristic feature of general solutions to the Wegner equation \labelcref{eq:WF-dotphi} and the integration constants are uniquely fixed by the required finiteness of $\dot \phi_t$ even in interacting theories.

Finally, we obtain a complex PIK given by 
\begin{align}
	\dot \phi_t =  -\frac{i}{m^2} \qquad \Rightarrow \qquad 
	\phi_t (\phi_0) = \phi_0 - \frac{i}{m^2}t \,,
\end{align}
where we integrated up initial samples $\phi_0 \sim p_0$ in the RG-time $t$ using \labelcref{eq:PhiMapFromCalc} in the second step as explained in the main text. The transformation corresponds to a shift of the field to the complex plane. The Jacobian of the map is given by $\partial_{\phi_0} \phi_t = 1$ which allows to easily verify \labelcref{eq:SignProblem}. This can also be seen by inserting $\phi_t(\phi)$ into the target action \labelcref{eq:ExFree_1}. Following \labelcref{eq:Idenity} this simply recovers the initial action
\begin{align}
	S_t(\, \phi_t(\phi_0)\, ) = \frac{1}{2} m^2 \phi_0^2 + \frac{t^2}{2 m^2} + \log \mathcal{N}_t = \frac{1}{2} m^2 \phi_0^2 + \log \mathcal{N}_0 \,,
\end{align}
i.e.~the action $S_t$ evaluated for fields that live on the manifold ${\cal M}_t$ -- which is parametrised by $\phi_t (\phi_0) $ -- is completely real for all $t$.

Finally, we can also compute correlation functions from this map: For example the two-point function evaluates to
\begin{align}
	\langle \phi_t^2\rangle &= \langle \,\left(\phi_0 - i t/m^2 \right)^2\,\rangle = \langle \phi_0^2 \rangle - 2 \frac{i}{m^2}t \langle  \phi_0 \rangle - \frac{t^2}{m^4} \langle 1 \rangle = \frac{1}{m^2} - 0 - \frac{t^2}{m^4} = \frac{m^2- t^2}{m^4} \,,
\end{align}
which matches the result of a direct evaluation of the integral. 

\begin{figure*}[t!]
	\centering
	\begin{subfigure}{.31\linewidth}
		\centering
		\includegraphics[width=\linewidth]{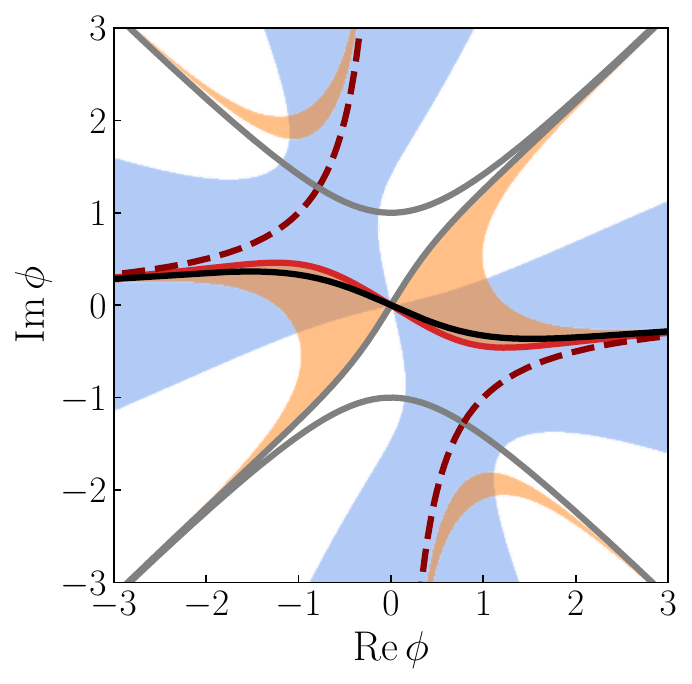}
		\subcaption{$ m^2 = 1 + 2 i$.}
		\label{fig:SymStar}
	\end{subfigure}%
	\hspace{0.02\linewidth}%
	\begin{subfigure}{.32\linewidth}
		\centering
		\includegraphics[width=\linewidth]{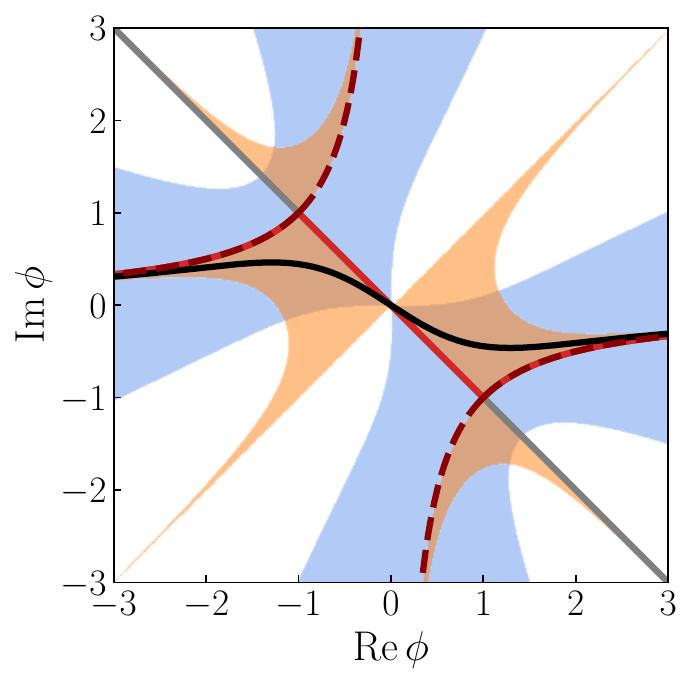}
		\subcaption{$ m^2 =  2 i$.}
		\label{fig:StokesStar}
	\end{subfigure}%
	\hspace{0.02\linewidth}%
	\begin{subfigure}{.32\linewidth}
		\centering
		\includegraphics[width=\linewidth]{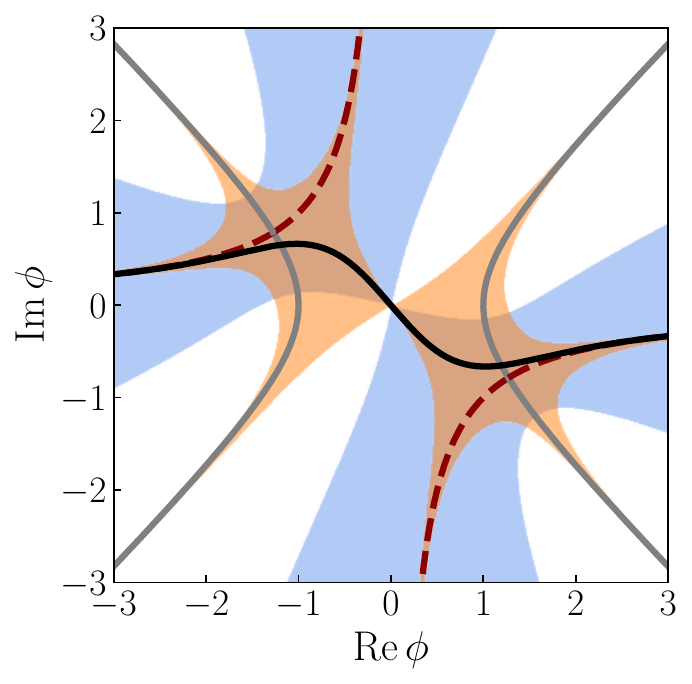}
		\subcaption{$ m^2 = -1 + 2 i$.}
		\label{fig:BrokenStar} 
	\end{subfigure}%
	\caption{PIKfold \textit{(black)} in the complex plane for the action \labelcref{eq:actionLT} with different values of $m^2$. In all cases we use $m^2_0 = 1$, $\lambda = 1$. We also depict the existing thimble solutions: The thimble with a real saddle at $\phi = 0$ \textit{(light red)}, the thimbles with complex saddles \textit{(dark red, dashed)} as well as the corresponding anti-thimbles \textit{(grey)}. Furthermore, the blue areas indicate $\Re[S(\phi)]>0$, i.e.~where the contour at infinity can be closed safely. The \textit{(orange)} areas indicate $0 \leq \Im[S(\phi)] \leq \frac{\pi}{2}$. \hspace*{\fill}}
	\label{fig:Stars}
\end{figure*}
%

\subsubsection{Complex $\mathbbm{Z}_2$-symmetric $\phi^4$-theory}
\label{app:ComplexZ2}

As pointed out previously, complex PIKs are related to other approaches for complex sampling, such as LTs and CLE. The complex $\mathbbm{Z}_2$-symmetric $\phi^4$-theory in particular is a good candidate for a comparison to the LT approach, as the thimbles of this model are well-studied~\cite{Aarts:2013fpa, Bharathkumar:2020kvz} and purely analytical for certain sets of couplings. Motivated by these solutions, we consider target actions with the following path
\begin{align}\label{eq:actionLT}
	\hat{S}_t(\phi) &= m_t^2 \frac{\phi^2}{2} + \lambda \frac{\phi^4}{4} \,, \qquad
	m_t^2 = m^2_0 + \Delta m^2  t \,,
\end{align}
where $m^2_0 = \lambda =1$ and different $\Delta m^2 \in \mathbbm{C}$. We consider three cases with (a) $\Delta m^2  = 2i$, (b) $\Delta m^2 = -1 +2i $ and (c) $\Delta m^2 = -2 +2i $. Accordingly, $m^2$ is shifted/rotated from the real axis into the complex plane: The path starts from a real action on $\mathcal{M}_0 = \mathbbm{R}$ at $t=0$ and is continuously deformed to some complex $\mathcal{M}_1$ at $t=1$. 
Note, that in this case, we do not modify the quartic coupling, i.e.~$\lambda$. This generally results in closed changes to the manifold $\mathcal{M}_t$, as discussed around \Cref{fig:ClosedDifference}.

The numerical procedure of solving the Wegner equation and computing the manifold is addressed in \Cref{app:PIKNumerics}.

The resulting sampling manifolds $\mathcal{M}_1$ (black) for the examples (a)-(c) are presented in \Cref{fig:Stars}.
The figures highlight regions where $\Re[ \hat{S}(\phi)]>0$ \textit{(blue)} and $|\Im[ \hat{S}(\phi)]\,| \leq \frac{\pi}{2}$ \textit{(orange)}. The blue region corresponds to the Stokes wedges, i.e.~where closing the contour is well-defined, which is the case for all examples. The orange regions indicate areas in which the phase of the distribution is only lightly varying and in particular not oscillating. This area is only meant as a visual guide, as the sampling on the manifold is by construction real due to the additional Jacobian, recall \labelcref{eq:dtWeightPT=0,eq:Idenity}.

To compare with the LT approach, we also depict the thimble solutions in \Cref{fig:Stars}. The thimble-structure differs quite drastically between the considered examples: (a) has one contributing thimble with a real saddle, (b) sits on a Stokes line and (c) has two contributing thimbles with complex saddles. By comparison, there is no change in the qualitative features of the PIKfolds, which have a similar shape across all the three examples. In particular the PIK is not identical to any thimble and does not reach the complex saddles. Nevertheless, it represents a closed contour and allows to quantitatively sample observables using \labelcref{eq:SampleOt}.

\subsubsection{Complex $\phi^4$-theory with a linear term}
\label{app:ComplexZ2+lin}

In the present section we provide further details for the computation of complex $\phi^4$-theories with a linear term, explicitly breaking the $\mathbbm{Z}_2$ symmetry of the $\phi^4$-theory. Firstly, we will discuss the example shown in the main text in \Cref{fig:Showcase}, where the coupling of the linear term is complex. Secondly, we discuss effects of using a complex mass term.

\textit{Complex linear term:} For convenience, we recall the action \labelcref{eq:Action0d}
\begin{align}
\hat{S}_t(\phi) = \frac12 \phi^2  +  \frac14\phi^4  + (1 + i t) \phi \,.
	\label{eq:Action0d_app}
\end{align}
The numerical details for the computation of the PIKs are deferred to \Cref{app:PIKNumerics}. In addition to the PIKs computed in the previous section, the missing $\mathbbm{Z}_2$ symmetry requires further basis functions in the numerical evaluation, see \labelcref{eq:basis}.

\begin{figure}[t]
	\centering
	\begin{subfigure}{0.45\textwidth}
		\centering
		\includegraphics[width=\linewidth]{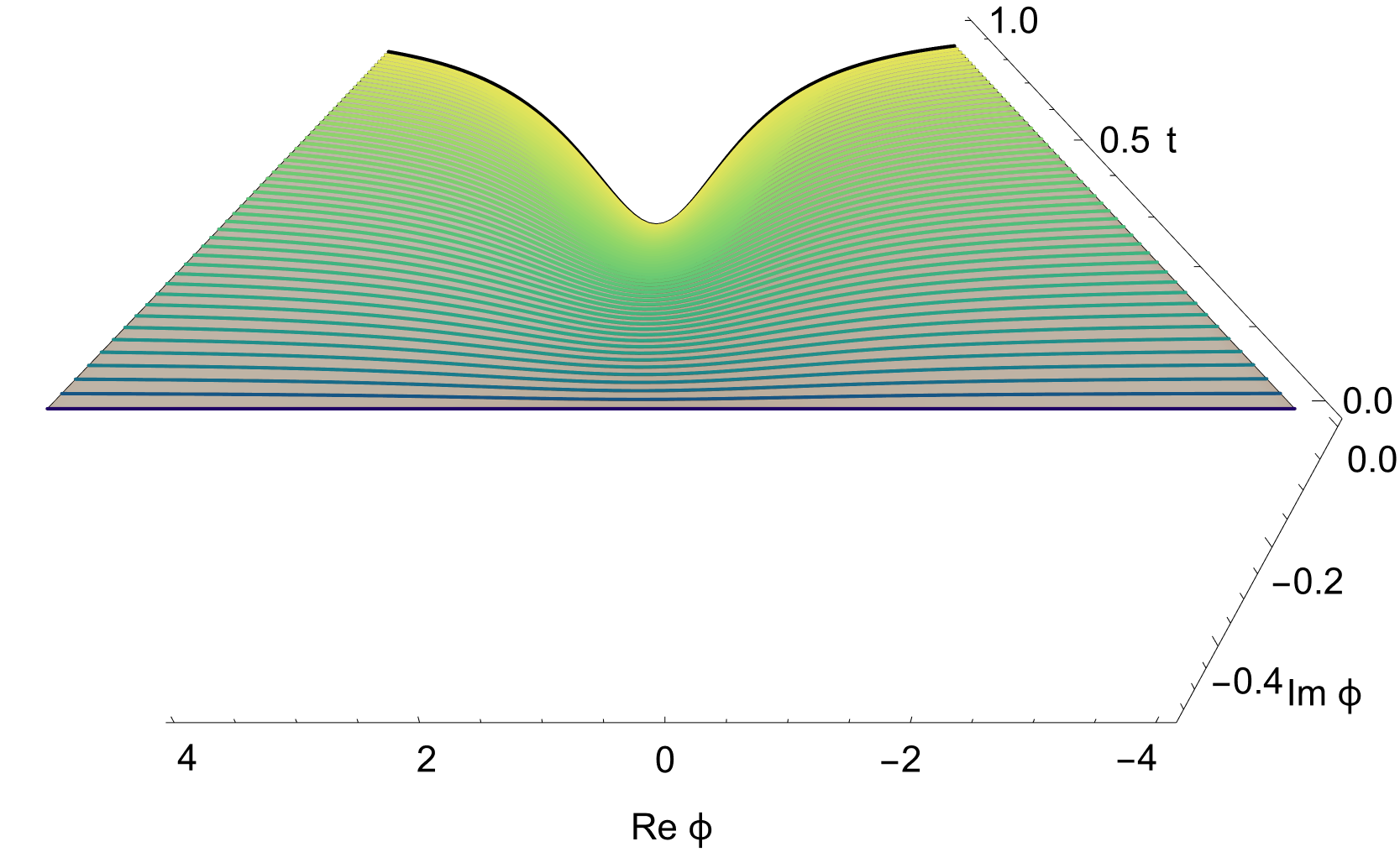} 
		\caption{PIKfold of the path \labelcref{eq:Action0d_app}.}
		\label{fig:PIKfoldMainText}
	\end{subfigure}%
	\hfill%
	\begin{subfigure}{0.45\textwidth}
		\centering
		\includegraphics[width=\linewidth]{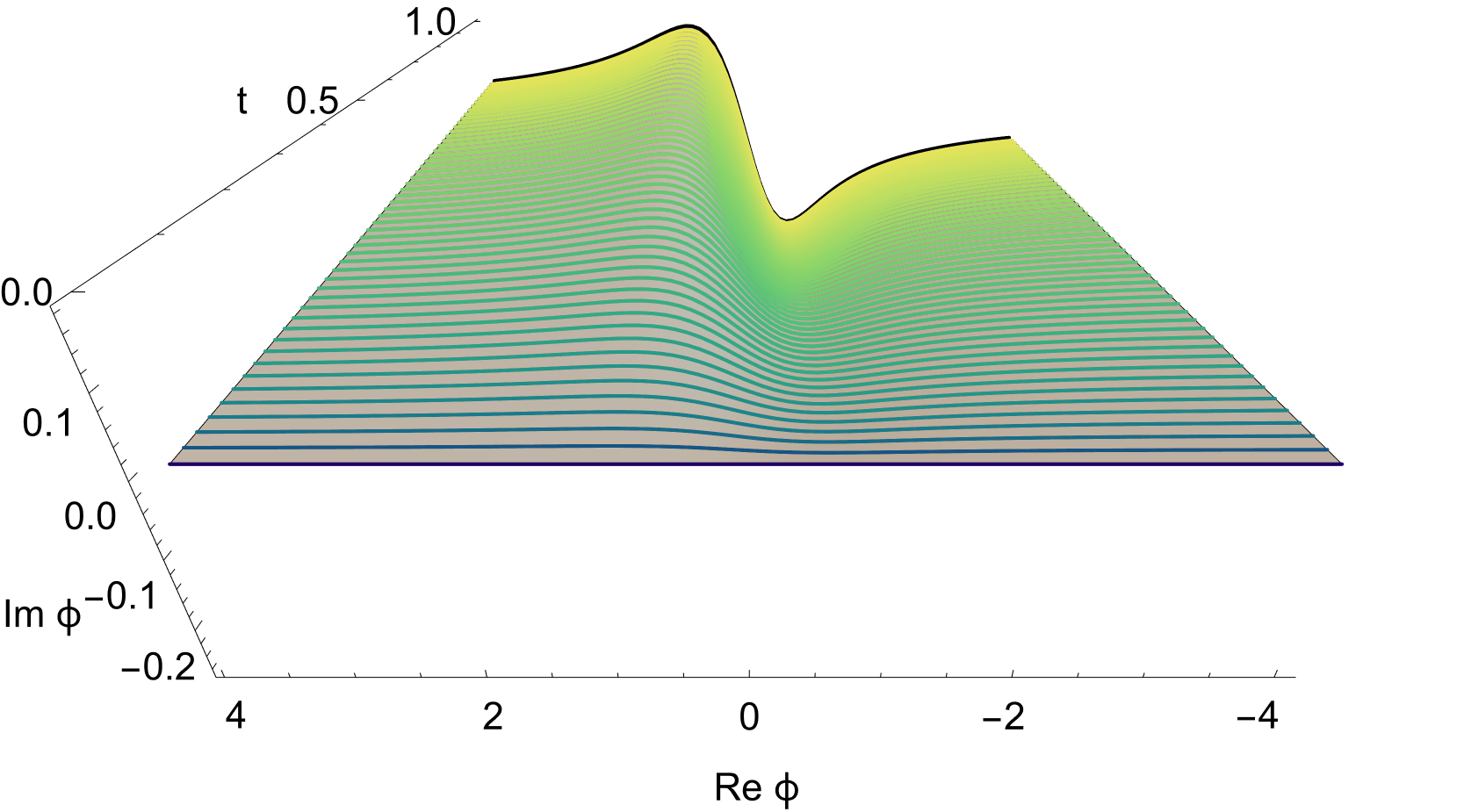}
		\caption{PIKfold of the path \labelcref{eq:Action0d_app1}.}
		\label{fig:PIKfoldAppendix}
	\end{subfigure}%
	\caption{Full PIKfolds $\mathcal{M}_t$ for the zero-dimensional examples with a linear term. The discrete time-slices that are used in the computation are highlighted in colour, ranging from the real axis $\mathcal{M} _0 = \mathbbm{R}$ to the final sampling manifold $\mathcal{M} _1$ which is coloured in black and depicted again in \Cref{fig:Showcase}.\hspace*{\fill}}

\end{figure}

The action \labelcref{eq:Action0d_app} is particularly challenging for the LT approach, as all thimbles have different residual phases. This leads to a global sign problem and multiple thimbles contribute to the evaluation of the integral~\cite{Aarts:2014nxa}. 
In comparison, the complex PIK approach computes a single sampling manifold $\mathcal{M}_1$, recall \Cref{fig:Showcase}.
We depict the full PIKfold $\{\mathcal{M}_t\}$ in \Cref{fig:PIKfoldMainText}, where one can clearly discern the successive build-up of the manifold from the real-axis at $t=0$ to the final sampling manifold $\mathcal{M}_1$.

\begin{figure*}[t!]
	\centering
	\begin{subfigure}{.31\linewidth}
		\centering
		\includegraphics[width=\linewidth]{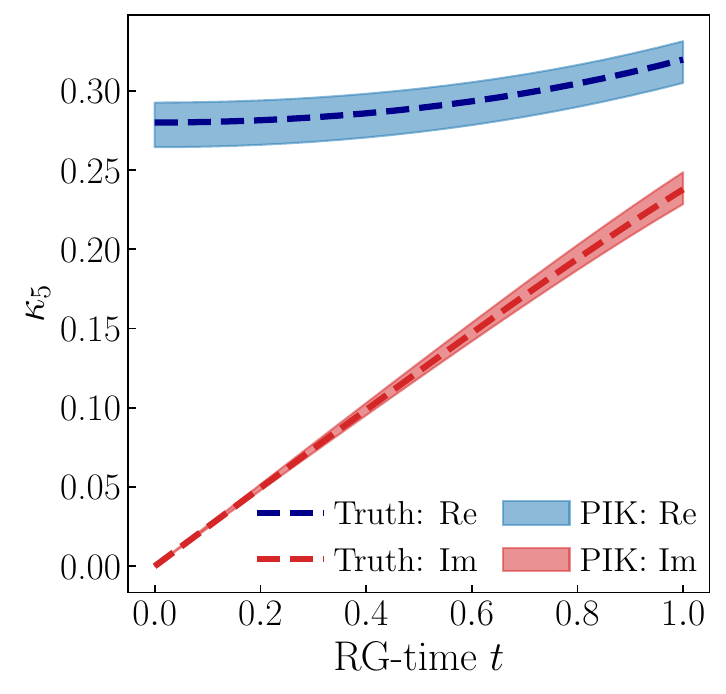}
		\subcaption{Evolution of the 5$^{\textrm{th}}$ cumulant.}
		\label{fig:Kappa5}
	\end{subfigure}%
	\hspace{0.01\linewidth}%
	\begin{subfigure}{.31\linewidth}
		\centering
		\includegraphics[width=\linewidth]{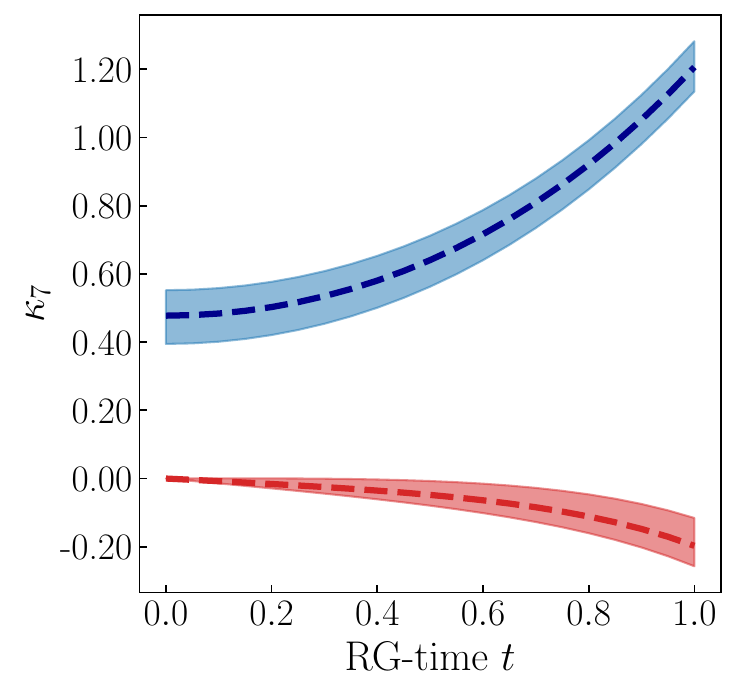}
		\subcaption{Evolution of the 7$^{\textrm{th}}$ cumulant.}
		\label{fig:Kappa7}
	\end{subfigure}%
	\hspace{0.01\linewidth}%
	\begin{subfigure}{.31\linewidth}
		\centering
		\includegraphics[width=\linewidth]{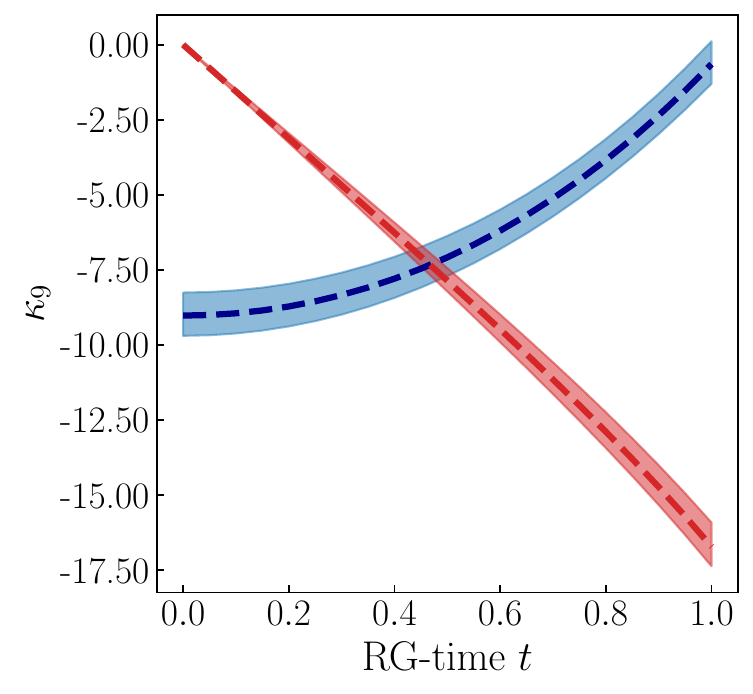}
		\subcaption{Evolution of the 9$^{\textrm{th}}$ cumulant.}
		\label{fig:Kappa9} 
	\end{subfigure}%
	\caption{RG-time dependent cumulants of the model with the action \labelcref{eq:Action0d_app} computed using $10^5$ Monte Carlo samples at $t=0$ and transporting them using \labelcref{eq:IntegrateMap} together with the physics-informed kernels for this model. As a visual guide we depict exact solutions computed by a direct numerical evaluation of the integral. \hspace*{\fill}}
	\label{fig:Observables_Phi4LinearTerm}
\end{figure*}

Finally, we compute the RG-time dependent cumulants $\kappa_n$ of the distributions given by the path $S_t$. To this aim, we first sample $10^5$ configurations from the purely real initial distribution at $t=0$ using a standard metropolis algorithm. Next, the real samples are evolved in RG-time with the PIKs, see \labelcref{eq:IntegrateMap}, and we evaluate the moments $\langle \phi_t^n \rangle$ at each time-step using \labelcref{eq:SampleOt}. Finally, we compute all cumulants up to $n=10$ using the moments. Results for the cumulants are shown in \Cref{fig:Observables_Phi4LinearTerm} where we depict the evolution of the fifth, seventh and ninth cumulant together with their statistical error.
Within this error, all cumulants are in agreement with the exact results obtained from a direct numerical evaluation of the respective integrals.

Note, that one could also have used a purely real PIK for the initial sampling of $S_0$. We have refrained from doing so, such that the error on the initial samples is that of a known, standard lattice simulation method. 

\begin{figure*}[t]
	\centering
	\begin{subfigure}{0.31\textwidth}
		\centering
		\includegraphics[width=\linewidth]{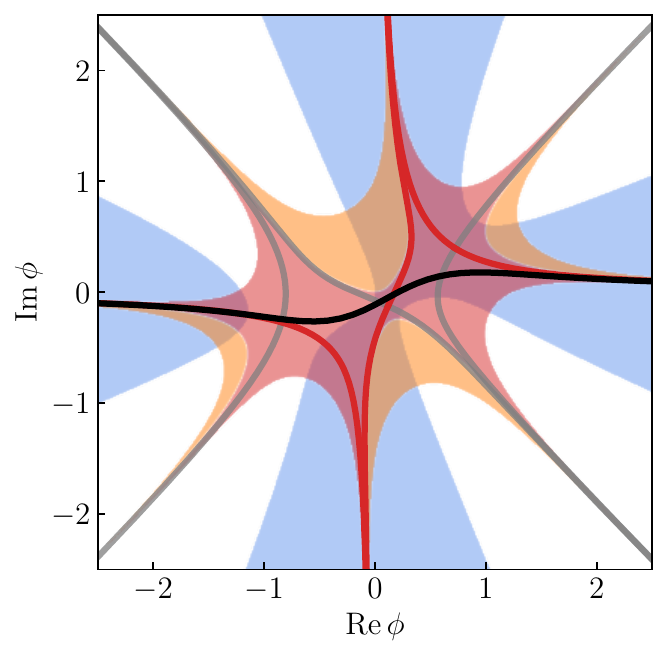}
		\caption{Sampling manifold $\mathcal{M}_1$. }
		\label{fig:PIKThimbleAppendix}
	\end{subfigure}
	\hfill
	\centering
	\begin{subfigure}{0.31\linewidth}
			\centering
			\includegraphics[width=\linewidth]{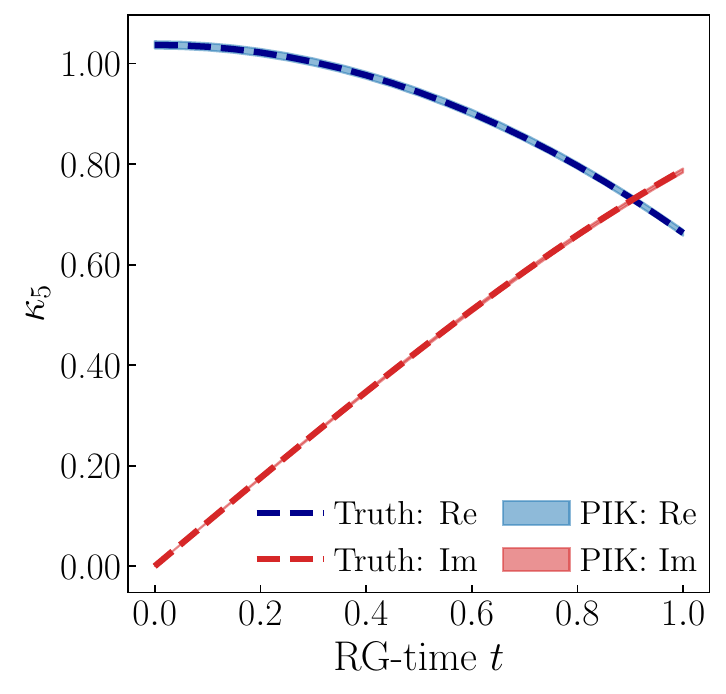}
			\subcaption{Evolution of the 5$^{\textrm{th}}$ cumulant.}
			\label{fig:Kappa5Appendix}
	\end{subfigure}
		\hfill
	\begin{subfigure}{0.31\linewidth}
			\centering
			\includegraphics[width=\linewidth]{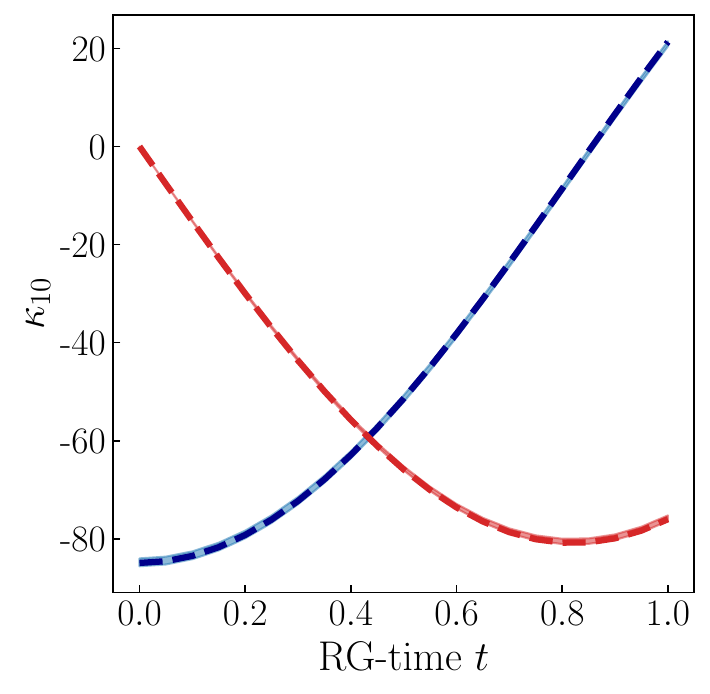}
			\subcaption{Evolution of the 10$^{\textrm{th}}$ cumulant.}
			\label{fig:Kappa10Appendix}
		\end{subfigure}
		\caption{Sampling manifold and cumulants for the action \labelcref{eq:Action0d_app1}. (a) The sampling manifold is shown in \textit{(black)}, (anti-)thimbles are shown in (grey) red. The shaded areas indicate: $\Re[S(\phi)]>0$ \textit{(blue)} and $0 \leq \pm \Im[S(\phi)] \leq \frac{\pi}{2}$ \textit{(orange/red)}. (b, c) RG-time dependent 5$^{\textrm{th}}$ and 10$^{\textrm{th}}$ cumulants computed from $10^5$ Monte Carlo samples at $t=0$. As a visual guide we also show exact solutions computed by a direct numerical evaluation of the integral. \hspace*{\fill}}
\end{figure*}

\textit{Complex mass term:} Now, following the same procedure, we also consider an example with a complex mass term. The action for this theory reads
\begin{align}\label{eq:Action0d_app1}
	\hat{S}_t(\phi) = -(1 + i t) \phi^2  +  \phi^4  + \frac12 \phi \,.
\end{align}
The full PIKfold is depicted in \Cref{fig:PIKfoldAppendix} and shows a two peak structure. The respective structure of the (anti)thimbles, and the final sampling contour are depicted in \Cref{fig:PIKThimbleAppendix}. Again, we observe that while the thimble structure with multiple contributing thimbles is quite complex, the single final sampling manifold remains simple. Also considering the cumulants, we depict the fifth and tenth cumulant in \Cref{fig:Kappa5Appendix,fig:Kappa10Appendix}. Here the statistical errors from the initial Monte Carlo approach are significantly smaller than in the previous example.

\subsubsection{Real-time proxy $\phi^4$-theory} 
\label{app:RealTimeProxy0d}

We now turn towards the simulation of real-time physics. As discussed in the main text around \labelcref{eq:RealTimePath}, one can simulate real-time physics with PIKs by rotating the real couplings of the lattice action at $t=0$ into purely imaginary ones at $t=1$. In the following, we first consider a free theory in zero-dimensions for which all computations can be done analytically to get an intuition for the setup. This is followed by the numerical evaluation of an interacting theory in zero-dimensions analogous to the discussion around \Cref{fig:Stars} in the main text. 

We consider the action of a zero-dimensional $\phi^4$-theory with RG-time dependent couplings
\begin{align}
	\hat{S}_t(\phi) =  m_t^2 \frac{ \phi^2}{2} +\lambda_t  \frac{\phi^4 }{4} \,, \quad \mathrm{with} \qquad 	m^2_t = m^2 \, e^{-i \frac{\pi}{2} t} \,, \quad \mathrm{and} \quad \lambda_t = \lambda \, e^{-i \frac{\pi}{2} t}\,,
	\label{eq:ZeroDimRealTimeAction}
\end{align}
and the time evolution of the parameters is chosen such that
\begin{align}
	\hat{S}_{t=0}(\phi) = \hat{S}(\phi) \quad \textrm{and} \quad \hat{S}_{t=1}(\phi) = -i \hat{S}(\phi) \,.
\end{align}
Which transforms the canonical Euclidean prefactor at $t=0$ into the real-time one at $t=1$, recall \labelcref{eq:prefactors}.  
This is done similarly to the prescription in the main text around \labelcref{eq:RealTimePath}. The difference in the sign in the exponent comes from the fact that we do not differentiate between a kinetic and potential term as the former is absent in zero-dimensions.

\textit{Free real-time theory.---} To begin with, we consider the free theory by setting $\lambda_t = 0$ as it allows for an analytical solution of the Wegner equation. In this explicit setting it reads
\begin{align}
	\frac{\partial \dot \phi(\phi)}{\partial \phi} - \dot \phi (\phi) \, m^2_t \phi = - \frac12 m^2_t\frac{ i \pi t}{2} \phi^2  + \frac{d\log \mathcal{N}_t}{dt}\,.
\end{align}
Again, there is only one finite solution, which is given by an appropriate choice of constants
\begin{align}\label{eq:Map2}
	\dot \phi =\frac{i \pi}{4} \phi \qquad \Rightarrow \qquad \phi_t(\phi_0) = \phi_0 \, e^{\frac{\pi}{4}it}\,,
\end{align}
with $\partial_t \log \mathcal{N}_t = \frac{\pi}{4}i $.
\Cref{eq:Map2} is an example for open changes of the manifold, as discussed around \Cref{fig:ClosedDifference}, which requires being able to close the integration contour at infinity. 

As in \Cref{app:FreeTheory0d}, in the second step, we perform the RG-time integration of initial samples $\phi_0 \sim p_0$ using \labelcref{eq:PhiMapFromCalc}. Given the field transformation $\phi_t(\phi_0)$, we can again compute the two-point function
\begin{align}
	\langle \phi_t^2 \rangle = e^{2 \frac{\pi}{4} it} \langle \phi_0^2\rangle =  e^{2 \frac{\pi}{4} it}\frac{1}{m^2} = \frac{1}{m^2 e^{-\frac{\pi}{2}i t}} \,,
\end{align}
where we inserted the correlation function of the real Gaussian distribution $p_0$ for $\langle \phi^2_0\rangle$.

The conservation of the overall weights \labelcref{eq:dtWeightPT=0} is also easily verified, by considering the pull-back to the original manifold
\begin{align}
	\hat{S}_t(\, \phi_t(\phi_0) \,) = \frac12 m^2 \phi_0^2 \,,
	\label{eq:ActionOnPIKfoldRealTime0d}
\end{align}
which simply evaluates to the original action $\hat S_0$, see \labelcref{eq:ZeroDimRealTimeAction}.

\Cref{eq:ActionOnPIKfoldRealTime0d} does not match $S_0$ in terms of the normalisation.
This discrepancy is explained by \labelcref{eq:Idenity} which relates the action $S_t$ for fields on the PIKfold to the initial action $S_0$, the Jacobian of the transformation and the normalisations. The latter two are given by
\begin{align}
	\log \mathcal{N}_t = -\frac{\pi}{4}i t  + \log \mathcal{N}_0\,, \qquad \log \left[\,\frac{\partial \phi_t}{\partial \phi_0}(\phi_0)\,\right] = \log e^{\frac{\pi}{4}it} = \frac{\pi}{4}it\,.
	\label{eq:NormalisationAndJacobianRealTime0D}
\end{align}
Finally, we compute \labelcref{eq:varphitApp} in this setup to gain some intuition for the appearing expressions. For this purpose, we first evaluate $\varphi_t$, which is given by
\begin{align}\label{eq:InverseMap}
	\varphi_t(\phi_t) = \phi_1 \qquad \Rightarrow \qquad \varphi_t(\phi) = \phi \, e^{-\frac{\pi}{4}i (t-1)} \,,
\end{align}
where we used knowledge of the map $\phi_1 = \phi_0  e^{\frac{\pi}{4}i}$ to rewrite \labelcref{eq:Map2}. Inserting this expression in \labelcref{eq:varphitApp} concludes this simple sanity check
\begin{align}
	\dot \phi(\phi)\,\frac{\partial \varphi^{\ }_t(\phi)}{\partial\phi} = \frac{i \pi}{4}\, \phi \,e^{-\frac{\pi}{4}i(t-1)} \qquad \Rightarrow \qquad \varphi^{\ }_t(\phi) = \phi - \int_1^t ds \,  \frac{i \pi}{4} \phi \, e^{-\frac{\pi}{4}i(s-1)} = \phi -  \phi  + \phi \, e^{-\frac{\pi}{4}i(t-1)} \,,
\end{align}
which is again \labelcref{eq:InverseMap}.  This concludes the analytical computations in the real-time free theory. \\[-2ex]

\begin{figure*}
	\centering
	\begin{subfigure}{.31\linewidth}
		\centering
		\includegraphics[width=\linewidth]{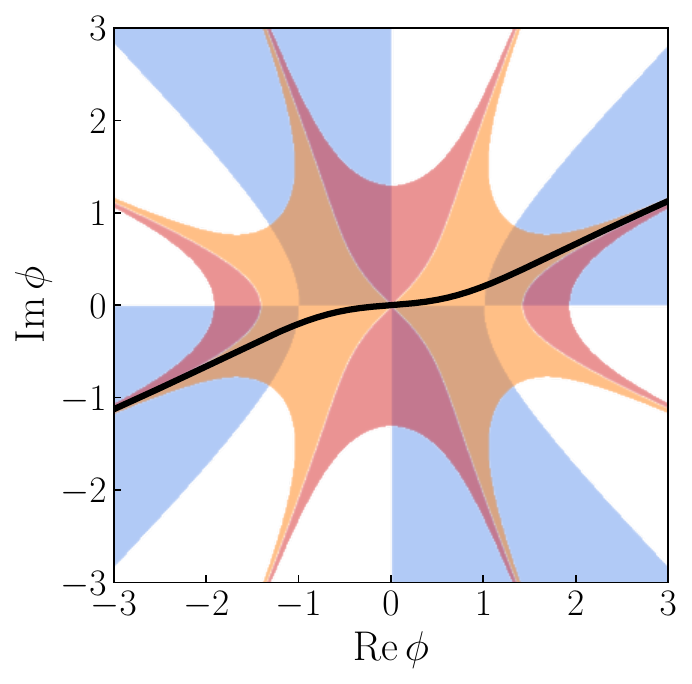}
		\subcaption{$m^2 = - 1$.}
		\label{fig:SymRStar}
	\end{subfigure}%
	\hspace{0.01\linewidth}%
	\begin{subfigure}{.31\linewidth}
		\centering
		\includegraphics[width=\linewidth]{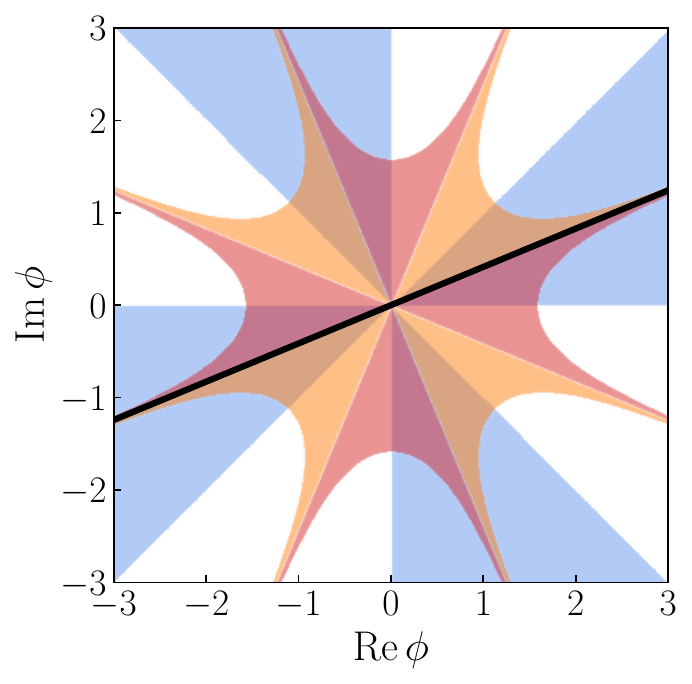}
		\subcaption{$m^2 = 0$.}
		\label{fig:NullStar}
	\end{subfigure}%
	\hspace{0.01\linewidth}%
	\begin{subfigure}{.31\linewidth}
		\centering
		\includegraphics[width=\linewidth]{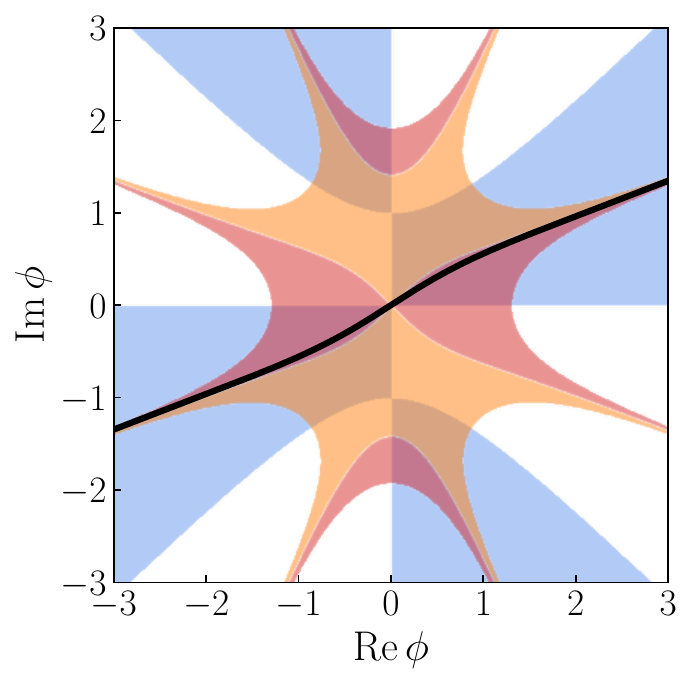}
		\subcaption{$m^2 =  1$.}
		\label{fig:BrokenRStar} 
	\end{subfigure}%
	\caption{PIK trajectory \textit{(black)} in the complex plane for different proxies of real-time actions.
		The blue areas indicate {$\Re[S(\phi)]>0$}, i.e.~where the contour at infinity can be safely closed. The \textit{(orange)} and \textit{(red)} areas indicate $0 \leq  \Im[S(\phi)] \leq \frac{\pi}{2}$ and $0 \leq - \Im[S(\phi)] \leq \frac{\pi}{2}$ respectively. In the latter two regions, both the real and imaginary part of the distribution can be sampled and the integrand is non-oscillatory.  \hspace*{\fill}}
	\label{fig:RealStars}
\end{figure*}

\textit{Interacting real-time theory.---} Having illustrated the central concepts of the complex PIK-architecture in the free theory, we now turn towards the interacting case. Here, we consider scenarios with positive, vanishing, and negative mass parameters $m^2$ and a non-vanishing quartic coupling $\lambda=1$ in the action \labelcref{eq:ZeroDimRealTimeAction}. The resulting PIK contour in the complex plane is shown in \Cref{fig:RealStars}. The numerical procedure to solve the Wegner equation in the interacting case is described in \Cref{app:PIKNumerics}. We find again, that the respective contours are located in areas with $|\Im[S(\phi)]|< \frac{\pi}{2}$ and allow for closing the integration contour at infinity.

\subsection{Free quantum field theories} 
\label{app:FreeTheoryAnyd}

In this section we discuss complex PIK-architectures in higher-dimensional theories. In particular, we consider the free scalar theory as a first, fully analytical benchmark. We detail the computation of the free complex PIK in arbitrary spacetime dimensions in \Cref{app:PIK-Gendim}. Secondly, in \Cref{app:PIK-freeQM}, we provide further details on the computation of the real-time quantum harmonic oscillator that was discussed in the main text \Cref{fig:FreeTheoryCorrelations}.

\subsubsection{Complex PIK-architecture for free scalar theories in general dimensions} 
\label{app:PIK-Gendim}

In this section, we discuss real-time simulations of a free scalar field theory in arbitrary spacetime dimensions using complex PIKs. To this aim we consider a lattice action given by
\begin{align}
	\hat{S}_t(\phi) = \sum_{i\in \mathcal{D}}\left[\,T_i(\phi) + V_i(\phi) \,\right] = \frac12 \sum_{i\in \mathcal{D}} \left[\,\sum_{\mu=0}^{d-1} \kappa_{t, \hat{\mu}} (\phi_{i+\hat{\mu}} - \phi_i)^2 + (m^2_t + \epsilon) \,\phi_i^2\,\right] = \frac12 \phi_i \, M_{t, i j} \, \phi_j \,,
	\label{eq:FreeActionAnyD}
\end{align}
where the matrix $M$ was introduced for convenience. The action contains the potential $V_i(\phi)$, with the mass parameters $m^2_t$ and $\epsilon$. While the parameter $m^2_t$ will be complexified during the flow with the PIKs, the small parameter $\epsilon$ is introduced to ensure the finiteness of the integral and remains constant. The kinetic contributions $T_i(\phi)$ of the action contain the parameters $\kappa_{t, \hat{\mu}}$, where $\hat{\mu}$ denotes the unit vector in the $\mu$-direction. Following the strategy of \Cref{app:RealTimeProxy0d}, the time-dependent couplings implement a rotation onto the imaginary axis, starting from a Euclidean signature at $t=0$ to a Minkowski signature at $t=1$. To this end we choose their time-dependence as 
\begin{align}
	\kappa_{t, \hat{\mu}} = e^{-i\kappa_{M, \mu} \frac{\pi}{2} t} \,, \quad m^2_t = m^2 e^{+i \frac{\pi}{2} t} \,,
\end{align}
with $\kappa_{M, 0} = 1$ and $\kappa_{M, i} = -1$ for $i=1,\ldots,d-1$. Finally, the normalisation of this theory in this notation is given by 
\begin{align}
	\mathcal{N}_t = \sqrt{\frac{(2\pi)^{|\mathcal{D}|}}{\det M_t}} \,,
	\label{eq:NormFreeAnyD}
\end{align}
where $|\mathcal{D}|$ denotes the number of lattice sites. With \labelcref{eq:NormFreeAnyD}, the change of the normalisation ${d} \mathcal{N}_t/dt$ can be computed explicitly using Jacobi's formula. Finally, the Wegner equation corresponding to this change of the action reads
\begin{align}
	\frac{\partial \dot{\phi}_{t,i}(\phi)}{\partial \phi_i} - \dot{\phi}_{t, i}(\phi) \, \, M_{t}{}^{i j} \, \phi_j = \frac12 \phi_i \frac{d  M_{t}{}^{i j} }{dt} \phi_j + \frac{d \log \mathcal{N}_t}{dt} \,.
	\label{eq:WegnerFreeAnyD}
\end{align}
Minding the required boundary conditions and symmetries, \labelcref{eq:WegnerFreeAnyD} is solved by
\begin{align}\label{eq:freePIK}
	\dot{\phi}_t(\phi) = - \frac12 M_t^{-1} \frac{d  M_t }{dt}\, \phi \,.
\end{align}
With access to this kernel, we can transform one free theory into another. Here, we note two aspects relevant to the generalisation to interacting theories. Firstly, the kernel contains the term $M_t^{-1}$ which is a non-sparse matrix, generally coupling all lattice sites to each other. This feature is also assumed to persist in more complicated theories. Consequently, also kernels for interacting theories must be able to capture these non-local terms in the solution to the Wegner equation. Secondly, we emphasise that the kind of non-locality appearing in the solution can be (approximately) known from the structure of the action. This provides a crucial benefit when computing the kernel in a more general setting and reflects the fact that the kernel is `physics-informed', i.e.~determined by some analytically known action and Wegner equation.

\subsubsection{Real-time harmonic oscillator} 
\label{app:PIK-freeQM}

\begin{figure}[t]
	\centering
	\includegraphics[width=0.5\textwidth]{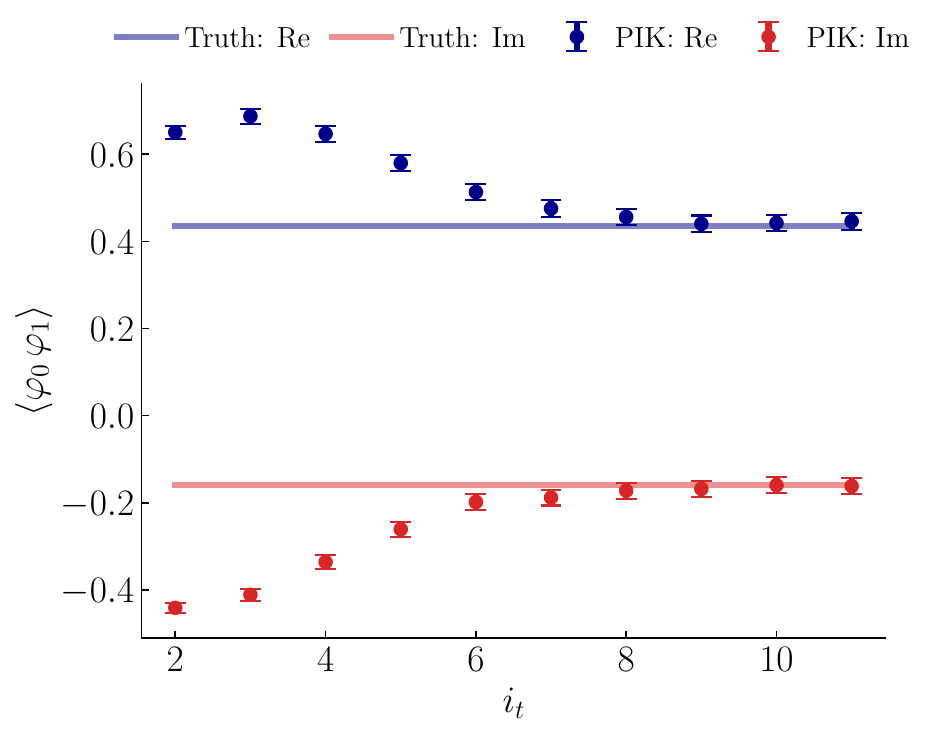}
	\caption{Correlation $\langle \varphi_0 \, \varphi_1 \rangle$ in the free theory \labelcref{eq:Action0d} for differently fine Euler time-discretisations of the RG-time evolution of fields \labelcref{eq:IntegrateMap} with $N_t = 2^{i_t}$ time-steps. The true solution is given by the solid line. The used parameters are $N_{\textrm{max}}=30$, $m^2=0.3$, $\epsilon = \frac{2m}{N_{\textrm{max}}}$. \hspace*{\fill}}
	\label{fig:TimeDiscretisationFreeTheory}
\end{figure}

We now turn towards the explicit example of the quantum harmonic oscillator in real-time, for which $d=1$. The kernels for this theory were given in the main text, recall \labelcref{eq:FreeActionAnyD,eq:freePIK}. As shown and discussed in the main text around \Cref{fig:FreeTheoryCorrelations}, the PIK-architecture allows us to compute the real-time correlation functions by transforming samples from the initial Euclidean distribution, see \labelcref{eq:IntegrateMap}. In principle, this transformation can be affected by two kinds of numerical errors. Firstly, the solution of the Wegner equation may be inaccurate, meaning that the computed kernel $\dot \phi_t$ does not solve the Wegner equation exactly. Secondly, the time-discretisation of the RG-time evolution of the fields may be too coarse, such that the numerical integration of \labelcref{eq:IntegrateMap} does not yield the correct result. In the present example, the solution of the Wegner equation is known analytically, evading the first source of numerical error. This allows us to quantify the second source of error. We show the difference between the exact solution and the numerical result for the correlation $\langle \varphi_0 \varphi_1 \rangle$ for differently fine Euler time-discretisations of the RG-time evolution in \Cref{fig:TimeDiscretisationFreeTheory}. We find that the numerical computation agrees with the exact solution when using $N_t \geq 2^9$ equidistant time-steps. Moreover, in our numerical experiments the required number of time-steps did not scale with the number of lattice sites when conducting the same computations from \Cref{fig:TimeDiscretisationFreeTheory} for lattices with $N_{\textrm{max}}$ ranging from $10$ to $70$. Together with more elaborate time-stepping routines, this suggests that the time-discretisation error of the PIK approach to the sign problem can be well-controlled.

We close this section with a brief discussion of the anharmonic oscillator, the quantum-mechanical $\phi^4$-theory. 
The sign problem of the real-time (an)harmonic oscillator and related systems has been tackled with a variety of methods, including CLE and LT, see e.g.~\cite{Alexandru:2017lqr,Takeda:2019idb,Joseph:2020gdh,Heinen:2022eyh,Woodward:2022pet,Boguslavski:2022dee,Lampl:2023xpb,Alvestad:2023jgl}. For an illustration of one of the key, distinguishing, properties of the PIK-architecture we consider the simulation of the anharmonic oscillator with CLE. Its standard formulation (without kernel) suffers from a convergence to the wrong solution due to boundary terms that arise from an insufficient decay of the distribution during the Langevin process~\cite{Lampl:2023xpb,Alvestad:2023jgl,Seiler:2023kes}. In the PIK-architecture, the flow of the distribution is computed directly. At each step it is in one-to-one correspondence with the physical one due to the weight-preserving property of the PIK. It has been checked that such a real PIKfold also exists for the anharmonic oscillator with a few sites. Moreover, it is suggestive that this property persists also for larger lattices. A full discussion of the anharmonic oscillator and related systems goes beyond the scope of the present work and shall be provided in \cite{PIK-QM2026}.
%

\subsection{Numerical approach to complex physics-informed kernels}
\label{app:PIKNumerics}

In this supplement we discuss the numerical tasks involved in the setup and computation of complex physics-informed kernels. As we will lay out below, the PIK-framework contain three numerical tasks that can be optimised to treat a given sign problem.

Firstly, one must identify a suitable initial sampling manifold $\mathcal{M}_0$. For the specific theories investigated in this work, the weight-preserving property of the PIKs leads to $\mathbbm{R}^\mathcal{D}$ being an optimal choice for $\mathcal{M}_0$ as it easily ensures an efficient sampling with real and positive weights. However, other theories may benefit significantly from different choices for $\mathcal{M}_0$ to reach a sign-problem free sampling manifold. This is the object of a forthcoming work~\cite{PIKF2026}.

Secondly, one must identify a path $S_t$ for the action that continuously connects an initial distribution, induced by $S_0$, to the targeted distribution, induced by $S_1$, see \labelcref{eq:GlobalMap}. Importantly, this path of the action influences the final sampling manifold $\mathcal{M}_1$ via the Wegner equation \labelcref{eq:WF-dotphi} and the transformation of the fields \labelcref{eq:IntegrateMap}. Hence, the optimisation of the path $S_t$ may be included in the optimisation of this manifold.

Thirdly, for a given initial manifold $\mathcal{M}_0$ and path $S_t$, the remaining numerical tasks lies in solving the Wegner equation \labelcref{eq:WF-dotphi} for the kernels $\dot \phi_t$ for all required time-slices. Doing so efficiently, especially for higher-dimensional theories, is part of ongoing work started in~\cite{Ihssen:2025ybn}. In the following, we discuss the numerical aspects of this task in more detail and comment on the extension to complex valued kernels.

As discussed in the main text around \labelcref{eq:IntegrateMap}, sampling on the sampling manifold $\mathcal{M}_1$ is achieved by transporting samples $\phi_{t=0}$ from the initial distribution $p_0$ on $\mathcal{M}_0$ along
\begin{align}
	\frac{d \phi_{t,i}}{dt} = \dot \phi_{t,i}(\phi_t) \,.
\end{align}
For the purposes of this work, we discretised this RG-time evolution using an Euler scheme. However, further time-stepping routines can be readily implemented, see~\cite{Ihssen:2025ybn}. Unless noted otherwise, we used $N_t=1000$ time-steps in the computations of this work when then samples were transported from $t=0$ to $t=1$. An exemplary discussion of the time-discretisation error is given in \Cref{app:PIK-freeQM}. This setup requires to solve the Wegner equation \labelcref{eq:WF-dotphi} for each time-slice. For the purpose of the following discussion the Wegner equation \labelcref{eq:WF-dotphi} is restated as
\begin{align}
	\left[ \frac{\partial}{\partial \phi_i} -  \frac{\partial \hat{S}_t(\phi) }{\partial \phi_i}\right]\, \dot{\phi}_{t,i}(\phi) =\frac{d \hat{S}_t(\phi)}{dt} + \frac{d \log \mathcal{N}_t}{dt} \,,
	\label{eq:WFdotphiNumerics}
\end{align}
where we have included the explicit dependence on the normalisation $\mathcal{N}_t$, using $S_t(\phi) = \hat{S}_t(\phi) + \log \mathcal{N}_t$.

In zero-dimensional cases, the Wegner equation \labelcref{eq:WFdotphiNumerics} corresponds to an ordinary differential equation, whose numerical solution is straight-forward. In higher dimensions, it is a first order linear partial differential equation (PDE) for the kernel $\dot{\phi_t}$. In the following, the most important numerical aspects of solving the Wegner equation are discussed in more detail. Firstly, \Cref{app:OtherNumerics} summarises important general aspects of the numerical evaluation of PIKs, such as determining the change of normalisation ${d}\mathcal{N}_t/dt $ and the implementation of boundary conditions. Finally, we describe the treatment of complex valued fields in \Cref{app:ComplexSpatial}.

\subsubsection{Treatment of the normalisation and boundary conditions} 
\label{app:OtherNumerics}

As discussed in \Cref{app:NormalisationReloaded}, the change of the normalisation $\mathcal{N}_t$ is given by the non-trivial expectation value
\begin{align}
	\frac{d \log \mathcal{N}_t}{dt}= - \int D\phi\, \frac{d\,\hat{S}_t(\phi)}{dt} \, p_t(\phi)  \,.
	\label{eq:DefChangeInNormalisation}
\end{align}
Actually computing (or approximating) this constant to the required precision -- especially in the presence of a sign problem -- is a complicated task which is preferably avoided in practise. One of the main advances in \cite{Ihssen:2025ybn}, is to circumvent the computation of ${d}\log \mathcal{N}_t/d t$ by considering differences of the Wegner equation, see \Cref{app:NormalisationReloaded} for more details. Then, the constant drops out. For this, one considers a reference configuration $\chi \in \mathcal{M}_t$ and solves
\begin{align}
	\left[ \frac{\partial}{\partial \phi_i} -  \frac{\partial \hat{S}_t(\phi)}{\partial \phi_i} \right]\, \dot{\phi}_{t,i}(\phi)
	- \left[ \frac{\partial}{\partial \phi_i} -  \frac{\partial \hat{S}_t(\phi)}{\partial \phi_i} \right]\, \dot{\phi}_{t,i}(\phi)\Big|_{\phi=\chi} = \frac{d \hat{S}_t(\phi)}{dt} - \frac{d \hat{S}_t(\chi)}{dt}\,,
	\label{eq:WFdotphiDiff}
\end{align}
which is still a linear first order PDE for $\dot{\phi}_t$. Another consequence, related to the normalisation is given by the required boundary conditions for the kernel. As shown for real theories in~\cite{Ihssen:2025ybn}, by means of a simple integration of the Wegner equation and the assumption of vanishing boundary terms, one finds with \labelcref{eq:DefChangeInNormalisation}, that the kernel has to satisfy
\begin{align}
	\int_{\partial \mathcal{M}_t} D\phi^i \, p_t(\phi) \, \dot{\phi}_{t,i}(\phi) = - \frac{d \log \mathcal{N}_t}{dt} - \int D\phi\, \frac{d\,\hat{S}_t(\phi)}{dt} \, p_t(\phi) \overset{\labelcref{eq:DefChangeInNormalisation}}{=} 0  \,.
	\label{eq:BoundaryCondition}
\end{align}
This requires a sufficient decay of the kernel $\dot{\phi}_t$ on the boundary $\partial \mathcal{M}_t$ of the manifold $\mathcal{M}_t$. Accordingly, with \labelcref{eq:WFdotphiDiff}, we have access to a tractable form of the Wegner equation and with \labelcref{eq:BoundaryCondition} we know the required boundary condition for the kernel. With this, we proceed with the concrete solution method to compute complex  $\dot{\phi}_t$ in practice.

\subsubsection{Parametrising the complex physics-informed kernel}
\label{app:ComplexSpatial}

Solving for the kernel corresponds to solving a first order linear PDE. As pointed out in~\cite{Ihssen:2025ybn}, the linear structure of the Wegner equation allows us to improve a given approximate solution of the Wegner equation by linearly adding corrections given by the current PDE-residual. While this is a crucial aspect of the PIK approach to sampling, we will not draw on this part here and refer the interested reader to~\cite{Ihssen:2025ybn}.

In practise, we consider complex degrees of freedom $\phi \in \mathcal{M}_t$, where $\mathcal{M}_t \subset \mathbbm{C}^\mathcal{D}$. Consequently, derivatives of the action $\hat{S}_t$ and kernel $\dot{\phi}_t$ are also complex. This is implemented by parametrising the kernel $\dot{\phi}_t$ in a suitable truncated holomorphic (or at least meromorphic) basis $\{K_{j,t}\}_{j=1}^M$ with $M$ elements and complex expansion coefficients $\{k_{j,t}\}_{j=1}^M$, such that
\begin{align}
	\dot{\phi}_t(\phi) = \sum_{j}^{M} k_{j,t} \,K_{j,t}(\phi) \,.
\end{align}
Once expressed in a basis, one must solve for the coefficients $k_{j,t}$ that minimise the residual of the Wegner equation in the form of \labelcref{eq:WFdotphiDiff}. For the computations in the main text, we evaluated the Wegner equation at $N$ collocation points $\phi^{(i)} \in \mathcal{M}_t$, which leads to a simple linear equation of the form
\begin{align}
	A_t \, k_t = b_t \,,
\end{align}
where the matrix $A_t$ and the vector $b_t$ are readily determined by the collocation points and the basis elements. Note that this procedure turns the problem of finding a good kernel into a linear problem rather than a non-linear one as is the case for many related approaches. One can additionally optimise the path of the action or improve shape parameters of the basis. This would often take the form of a non-linear optimisation. However, the indispensable computational task of computing the kernel can always be written as a linear problem.

The efficiency of this linear problem depends crucially on the choice of basis functions. Therefore, we will elaborate briefly on the choice of basis elements for the chosen $\phi^4$-theories. To begin with, the number of coefficients can be reduced by considering symmetries. For the quantum-mechanical $\phi^4$-theories considered here, this includes translational symmetry, isotropy, and a global $\mathbbm{Z}_2$-symmetry. As is known from the literature on generative models~\cite{Koehler:2020efe}, to ensure that the symmetry also holds for the modelled action, the kernel should transform equivariantly under these symmetry transformations. In particular, the $\mathbbm{Z}_2$-symmetry requires that the kernel is an odd vector field with respect to the fields, i.e.
\begin{align}
	\dot{\phi}_t(-\phi) = - \dot{\phi}_t(\phi) \,.
\end{align}
This leads to the fact that $\dot{\phi}_t(0) = 0$ and respectively, $0 \in \mathcal{M}_t$ for all $t$. Here, we use this fact and conveniently choose $\chi = 0$ for all time steps. Moreover, we only choose kernel basis elements that are odd in the fields. Now, we go on and consider the behaviour of the Wegner equation \labelcref{eq:WFdotphiNumerics} together with the boundary condition \labelcref{eq:BoundaryCondition}. Whereas the right-hand side of \labelcref{eq:WFdotphiNumerics} is dominated by the $d S_t(\phi)/dt$ term which grows with the fourth power of the fields, the left-hand side contains the $\partial_\phi S_t(\phi)$ term that grows with the third power of the fields. Together with the boundary condition, this suggests choosing kernel basis elements that grow linearly with the field values. This is readily achieved by a rational function basis. For the zero-dimensional case, we considered basis elements of the form
\begin{align}
	K_{j, t}(\phi) = \frac{\phi^{n_j}}{( 1 + \phi^2)^{m_j} } \, \quad \textrm{with} \quad n_j \leq 2 m_j + 1 \,, \quad n_j \textrm{ odd} \,,
\label{eq:basis}
\end{align}
where one considers all possible combinations of $n_j$ and $m_j$ up to some maximal value of $m_{\textrm{max}}$. For the examples depicted in \Cref{fig:Stars,fig:RealStars}, using $m_{\textrm{max}} = 25$ leads in most cases to an average residual norm of the order of $10^{-6}$ and a relative residual norm of the order of $10^{-9}$. Here we define the average relative residual norm as $\| A_t k_t - b_t \|_2 \,/\, \| b_t \|_2$. This concludes the discussion of the numerical approach to complex physics-informed kernels which is part of actively ongoing research.

\end{document}